\documentclass{elsarticle}
\usepackage[utf8]{inputenc}
\usepackage{graphicx}
\usepackage{comment}
\usepackage{natbib}
\usepackage{amsmath}
\usepackage{amssymb}
\usepackage{lineno, comment, url, textcomp}
\usepackage[usenames,dvipsnames,svgnames,table]{xcolor}

\usepackage[english]{babel}

\newcommand{\TP}[1]{\mathcal{P}\left(#1\right)}

\newcommand{\rr}{\mathbf{r}}

\newcommand{\diff}{\,\mathrm{d}}
\newcommand{\IT}{\mathbf{I}}
\newcommand{\uu}{\mathbf{u}}

\newcommand{\LL}{\mathbf{L}}

\newcommand{\xx}{\mathbf{x}}

\newcommand{\pp}{\mathbf{p}}

\newcommand{\eps}{\varepsilon}

\newcommand{\yy}{\mathbf{y}}

\newcommand{\ra}{\rho_\alpha}

\newcommand{\jj}{\mathbf{j}}
\newcommand{\kk}{\mathbf{k}}

\newcommand{\nn}{\mathbf{n}}

\newcommand{\fN}{{f_N}}
\newcommand{\fNS}{{f_N^s}}
\newcommand{\fns}{{f_n^s}}
\newcommand{\intd}[1]{\int\!\!\diff #1}
\newcommand{\drr}{\diff\rr}
\newcommand{\dpp}{\diff\pp}
\newcommand{\OO}{\mathbf{1}}
\newcommand{\TT}{\mathbf{2}}
\newcommand{\TTT}{\mathbf{3}}
\newcommand{\NN}{\mathbf{N}}
\newcommand{\nnn}{\mathbf{n}}
\newcommand{\aaa}{\mathbf{a}}
\newcommand{\bb}{\mathbf{b}}
\newcommand{\ii}{\mathbf{i}}
\newcommand{\dO}{\diff \OO}
\newcommand{\dN}{\diff \NN}

\newcommand{\PB}[3]{\left\{#1,#2\right\}^{(#3)}}
\newcommand{\iaia}{\mathbf{i_a}}
\newcommand{\ibib}{\mathbf{i_b}}
\newcommand{\sff}{s_f}
\newcommand{\trho}{\tilde{\rho}}
\newcommand{\tuu}{\tilde{\uu}}
\newcommand{\ts}{\tilde{s}}
\newcommand{\fNN}{f_{N\tilde{N}}}
\newcommand{\tOO}{\tilde{\OO}}
\newcommand{\tNN}{\tilde{\NN}}
\newcommand{\tf}{\tilde{f}}
\newcommand{\ti}{\tilde{i}}
\newcommand{\tii}{\tilde{\ii}}
\newcommand{\taa}{\tilde{\aaa}}
\newcommand{\tbb}{\tilde{\bb}}
\newcommand{\tfN}{\tilde{\fN}}
\newcommand{\trr}{\tilde{\rr}}
\newcommand{\tpp}{\tilde{\pp}}
\newcommand{\tu}{\tilde{u}}
\newcommand{\tsf}{\tilde{s_f}}
\newcommand{\tp}{\tilde{p}}
\newcommand{\sT}{s^T}
\newcommand{\uT}{u^T}
\newcommand{\uuT}{{\uu}^T}

\newcommand{\cref}{c_0}

\newcommand{\stg}{{\ensuremath{-\kern-4pt{\ominus}\kern-4pt-}}}

\newcommand{\FFF}{\mathcal{F}}
\newcommand{\XX}{\mathbf{X}}

\newcommand{\fPiN}{f_N}

\newcommand{\RR}{\mathbf{R}}
\newcommand{\PP}{\mathbf{P}}
\newcommand{\en}{e_n}
\newcommand{\ei}{e_i}
\newcommand{\ej}{e_j}
\newcommand{\fn}{f_n}
\newcommand{\rhoh}{\hat{\rho}}
\newcommand{\bbb}{\mathbf{b}}
\newcommand{\cc}{\mathbf{c}}
\newcommand{\ww}{\mathbf{w}}

\newtheorem{theorem}{Theorem}

\newtheorem{remark}[theorem]{Remark}

\newcommand{\uL}{{^\uparrow L}}
\newcommand{\uLL}{{^\uparrow \mathbf{L}}}
\newcommand{\dLL}{{^\downarrow \mathbf{L}}}
\newcommand{\dL}{{^\downarrow L}}

\begin{document}

\title{A hierarchy of Poisson brackets in non-equilibrium thermodynamics}
\author[ec,ntc,mff]{Michal Pavelka\corref{cor1}}
\ead{michal.pavelka@email.cz}
\author[ntc,fjfi]{Václav Klika}
\author[gebze]{O\u{g}ul Esen}
\author[ec]{Miroslav Grmela}

\cortext[cor1]{Corresponding author}

\address[ec]{École Polytechnique de Montréal, C.P.6079 suc. Centre-ville, Montréal, H3C 3A7, Québec, Canada}
\address[ntc]{New Technologies - Research Centre, University of West Bohemia,  Univerzitn\'{i} 8, 306 14 Pilsen, Czech Republic}
\address[mff]{Mathematical Institute, Faculty of Mathematics and Physics, Charles University in Prague, Sokolovská 83, 186 75 Prague, Czech Republic}
\address[fjfi]{Department of Mathematics, FNSPE, Czech Technical University in Prague, Trojanova 13, Prague 2, 120 00, Czech Republic}
\address[gebze]{Department of Mathematics, Gebze Technical University, Gebze-Kocaeli 41400, Turkey}

\begin{abstract}


Reversible evolution of macroscopic and mesoscopic systems can be conveniently constructed from two ingredients: an energy functional and a Poisson bracket. The goal of this paper is to elucidate how the Poisson brackets can be constructed and what additional features we also gain by the construction. In particular, the Poisson brackets governing reversible evolution in one-particle kinetic theory, kinetic theory of binary mixtures, binary fluid mixtures, classical irreversible thermodynamics and classical hydrodynamics are derived from Liouville equation. 
Although the construction is quite natural, a few examples where it does not work are included (e.g. the BBGKY hierarchy). Finally, a new infinite grand-canonical hierarchy of Poisson brackets is proposed, which leads to Poisson brackets expressing non-local phenomena such as turbulent motion or evolution of polymeric fluids. Eventually, Lie-Poisson structures standing behind some of the brackets are identified.

\textit{Accepted in Physica D: Nonlinear phenomena.\\
DOI 10.1016/j.physd.2016.06.011.\\
\textcopyright This manuscript version is made available under the CC-BY-NC-ND 4.0 license \url{http://creativecommons.org/licenses/by-nc-nd/4.0/}}

\end{abstract}

\begin{keyword}
Poisson bracket \sep hierarchy \sep projection \sep non-equilibrium thermodynamics \sep GENERIC \sep Lie-Poisson equation.

\PACS 05.70.Ln \sep 05.90.+m
\end{keyword}

\maketitle

\numberwithin{equation}{section}

\section{Introduction}
The principal goal of statistical mechanics is to extract pertinent features from the complete microscopic information contained in trajectories of microscopic particles composing  macroscopic systems under consideration. The pertinent features are those seen in mesoscopic and macroscopic experimental observations. We look for a reduced (mesoscopic) time evolution which concentrates only on the pertinent features and ignores the irrelevant details. The reduction process from the micro to the meso dynamics (consisting of three steps, (i) finding microscopic trajectories, (ii) extracting from them the pertinent pattern, and (iii) constructing mesoscopic dynamics whose trajectories reproduce the pattern) is, in general, very complex. It is therefore  very useful to realize that the microscopic, the mesoscopic,  and the macroscopic time evolutions share a common structure, namely that the reversible evolution is expressed by means of a Poisson bracket and energy, see \cite{DV, G1, Mor, Kauf, Miroslav-PhysicaD, Grmela1997, Ottinger1997}.

The two latter references introduce the General Equation for Non-Equilibrium Reversible-Irreversible Coupling (GENERIC). The GENERIC framework contains a large amount of mesoscopic models, e.g. classical hydrodynamics, Boltzmann equation, classical irreversible thermodynamics (CIT), extended irreversible thermodynamics (EIT), see \cite{Lebon-Understanding}, models for polymer flows, visco-elasto-plastic solids \cite{Hutter-plastic}, etc. The main features of GENERIC are that the reversible part of the evolution equations is constructed from a Poisson bracket while the irreversible from a dissipation potential (or dissipative bracket when thermodynamic forces are small), and that the equations are also automatically compatible with equilibrium thermodynamics as they gradually approach equilibrium. The Hamiltonian structure of reversible evolution is common to a large class of models in continuum physics on many levels of description, see also \cite{dissertation}, and we shall thus require that passages from levels with more details to levels with less details preserve the structure.

In this paper we focus on relations between Hamiltonian structures governing reversible evolution on different levels of description. This means that on the level of the  microscopic dynamics of all particles composing macroscopic systems we consider the complete dynamics but on mesoscopic levels we consider only a part of the  dynamics - the reversible part. For example, on the level of fluid mechanics,  we consider only the Euler part and on the level of kinetic theory only the free flow  without collisions. Moreover, we investigate only the Poisson brackets of the Hamiltonian structures, not the energy functionals.  Our goal is to construct  Poisson brackets as reductions of a known Poisson bracket expressing kinematics  on a more detailed (more microscopic) level.

For example, we can start on the Liouville level, i.e. on the level  on which  N-particle distribution functions serve as state variables. The time evolution on the Liouville level, governed by the Liouville equation  (see e.g. \cite{Zwanzig}), is Hamiltonian and reversible\footnote{The reversibility can be checked the same way as for the reversible part of Boltzmann equation in \cite{PRE2014} or it can be seen from compatibility of Liouville equation and Hamilton canonical equations.} time evolution. The right hand side of the Liouville equation  (Liouville vector field)  is gradient of energy transformed into a vector by the Liouville-Poisson structure (expressed mathematically in the Liouville-Poisson bracket - see Sec. \ref{sec.PB.Liouville}).  Projections  from the N-particle distributions to  lower (less detailed) levels of description then induce projections of the Liouville Poisson bracket to   Poisson brackets on the lower levels. The method of projection that we employ is quite natural and was already used for example in \cite{Marsden-BBGKY}, \cite{Miroslav-PhysicaD},    \cite{Ottinger} and in \cite{Elafif1999}. In this paper we present it   in its full generality within the context of non-equilibrium thermodynamics and investigate its geometric interpretation in appendices and Electronic supplementary (ES). We also  employ it to develop new levels suitable for applications in the theory of turbulence.
Various reductions demonstrated in this paper are presented in Fig. \ref{fig.map}.

It sometimes happens that reversible evolution is not given by projection of the Liouville Poisson bracket, as for example in an EIT theory of mixtures and BBGKY hierarchy (see Sec. \ref{sec.out}). What is then the common structure shared by the reversible evolution on different levels of description? In general this question still remains an open problem in non-equilibrium thermodynamics as it is not for example clear whether validity of Jacobi identity is necessary when the reversible evolution is accompanied by an irreversible counterpart. In particular examples, for example the BBGKY hierarchy, the common structure is again the Poisson bracket as the terms containing higher-order distribution functions can be replaced by a constitutive relation (closure) while the rest remains Hamiltonian. That is how the Boltzmann collision term comes into play while the reversible part of Boltzmann equation is Hamiltonian. The BBGKY hierarchy however is not part of the present hierarchy of Poisson brackets. It is thus still unclear whether the Hamiltonian structure should be a necessary feature of all consistent reversible dynamics describing real physical systems compatible with non-equilibrium thermodynamics or not, but even if it is, the Poisson bracket can be out of the present hierarchy of Poisson brackets.

When starting from the N-particle Liouville Poisson bracket, the number of particles is inherently incorporated into the Poisson brackets. To remove this dependence, we develop a grand-canonical hierarchy of Poisson brackets in Sec. \ref{sec.GCE}. This hierarchy can be then reduced to a two-point kinetic theory or to a Poisson bracket coupling hydrodynamic fields with a conformation tensor, Reynolds stress and non-local vorticity. Such an extended hydrodynamics should be useful in modeling of fluids where non-local effect play an important flow, e.g. in turbulence. 

Contribution of this paper can be seen in the following points. Firstly, the natural hierarchy of Poisson brackets generated by projections from N-particle distribution functions to less detailed levels of description is investigated in the context of non-equilibrium thermodynamics. Various thermodynamically relevant models are shown to be part of the hierarchy. Some of the projections seem to be novel (e.g. the passages to mixtures or to the symmetric distribution functions).
This systematic treatment of Poisson brackets should serve as a (so far lacking) guide for constructing Poisson brackets within the theory of non-equilibrium thermodynamics.
Secondly, a new hierarchy of non-local Poisson brackets is suggested, namely the grand-canonical ensemble hierarchy. This hierarchy leads to a two-point kinetic theory and to a Poisson bracket of extended hydrodynamics with applications in turbulent fluid motions, polymeric fluids, colloidal systems and other mesoscopic systems.
Thirdly, some of the projections are rewritten as the duals of some certain Lie algebra homomorphisms in the appendices, which should be interesting from the geometrical point of view.

\section{Hierarchy of Poisson brackets}\label{sec.PB}

\subsection{Motivation through introductory examples}
The most detailed description of a physical system seen as composed of classical particles is given by Hamilton's canonical equations or equivalently, in the Liouville representation, by the Liouville equation, which describes the time evolution of the N-particle distribution function. The Liouville level of description, where N-particle distribution function is the state variable, is the highest (most detailed) level of description and we thus start on this level.

The Hamiltonian structure of Liouville equation is presented in Sec. \ref{sec.PB.Liouville}. Then we shall observe in Sec. \ref{PB.sec.B} how the Poisson bracket generating the reversible part of Boltzmann equation can be obtained from the Poisson bracket generating Liouville equation by projecting the N-particle distribution function to a one-particle distribution function. This passage is then interpreted geometrically in Sec. \ref{PB.sec.geoint}, where also additional features of the passage pertinent to non-equilibrium thermodynamics are presented. The passage can be also seen as a Lie-Poisson reduction, see \ref{sec.LB.groups}.

\subsubsection{Hamiltonian structure of Liouville equation}\label{sec.PB.Liouville}
Consider an isolated system consisting of  $N$ interacting particles. The corresponding phase space consists of positions and momenta of all particles. Position and momentum of the first particle is denoted by $\rr_1$ and $\pp^1$, respectively, and analogously for other particles. The couple $(\rr_1,\pp^1)$ is denoted by $\OO$ for convenience and analogically for the other particles. Evolution equation of the N-particle phase space density $\fN(\OO,\dots,\NN, t)$, or N-particle distribution function, is the standard Liouville equation, see e.g. \cite{Gibbs-Liouville,Zwanzig},
\begin{equation}\label{PB.eq.Liouville}
 \frac{\partial \fN}{\partial t} = -\frac{\partial h_N}{\partial p^i_\alpha} \frac{\partial \fN}{\partial r^\alpha_i} + \frac{\partial h_N}{\partial r^\alpha_i}\frac{\partial \fN}{\partial p^i_\alpha}
\end{equation}
where the N-particle Hamiltonian $h_N(\OO,\dots,\NN)$ is a real-valued function of the position coordinates and momenta, that has the physical interpretation of the energy of the system. The Liouville equation is usually derived within non-equilibrium statistical physics from the requirement that probability density behaves as an incompressible ideal fluid. Note, however, that there is also an elegant geometric derivation exploiting properties of the coadjoint action of the group of canonical transformations on the cotangent bundle of classical mechanics, see \ref{sec.LB.groups}.

Energy on the Liouville level, $E$, can be expressed as
\begin{equation}
 E = \intd{\OO}\dots \intd{\NN} \fN h_N.
\end{equation}
Since
\begin{equation}
 \frac{\partial E}{\partial \fN} = h_N,
\end{equation}
Eq.\eqref{PB.eq.Liouville} implies that an arbitrary functional $A(\fN)$ of the distribution function  evolves as
\begin{eqnarray}
 \frac{\diff A}{\diff t} =\left\langle \frac{\partial A}{\partial \fN}, \frac{\partial \fN}{\partial t}\right\rangle= \int\dO\dots\int\dN\frac{\partial A}{\partial \fN}\frac{\partial \fN}{\partial t} = \PB{A}{E}{L}
\end{eqnarray}
where the Poisson bracket on the Liouville level (or the Liouville-Poisson bracket) is equal to
\begin{equation}\label{PB.eq.PB.Liouville}
 \PB{A}{B}{L} = \int\dO\dots\int\dN \fN\left(\frac{\partial A_\fN}{\partial r^\alpha_i}\frac{\partial B_\fN}{\partial p^i_\alpha}-\frac{\partial B_\fN}{\partial r^\alpha_i}\frac{\partial A_\fN}{\partial p^i_\alpha}\right)
\end{equation}
Note the usage of Einstein's summation convention over matching upper and lower indexes.
Integration with respect to for example $\dO$  stands for $\int \drr_1\int\dpp^1$, etc. This Poisson bracket can also be  rewritten in terms of the Poisson bracket of classical mechanics as
\begin{equation}\label{eq.Liouville.PB}
 \PB{A}{B}{L} = \int\dO\dots\int\dN \fN\PB{A_\fN}{B_\fN}{CM}
\end{equation}
where the Poisson bracket of classical mechanics is the standard canonical Poisson bracket, see e.g. \cite{Goldstein},
\begin{equation}\label{PB.eq.PB.CM}
 \PB{h_N}{k_N}{CM} = \frac{\partial h_N}{\partial r^\alpha_i}\frac{\partial k_N}{\partial p^i_\alpha}-\frac{\partial k_N}{\partial r^\alpha_i}\frac{\partial h_N}{\partial p^i_\alpha}
\end{equation}
for any two functions $h_N(\OO,\dots,\NN)$ and $k_N(\OO,\dots,\NN)$. The derivative $A_\fN$ stands for functional derivative of $A$ with respect to $\fN$ and will be denoted also by $\frac{\partial A}{\partial \fN}$.

Liouville equation can be thus constructed from the Liouville-Poisson bracket and energy $E(\fN)$. The Liouville-Poisson bracket will serve as a starting point for derivations of less detailed Poisson brackets.


\subsubsection{From Liouville to Boltzmann}\label{PB.sec.B}
Boltzmann equation consists of both reversible and irreversible dynamics. Let us now derive the reversible part from the Liouville-Poisson bracket, \eqref{PB.eq.PB.Liouville}. One-particle distribution function (or one-particle phase space density) $f(\rr, \pp)$, evolution of which is described by Boltzmann equation, can be defined as a projection of the N-particle distribution function. Indeed when denoting $(\rr_a, \pp^a)$ by $\aaa$, the one-particle distribution function can be introduced as
\begin{equation}\label{PB.eq.ffN}
 f(\aaa) = \int\dO\dots\int\dN \fN\sum_{i=1}^N\delta(\aaa-\ii)
\end{equation}
where $\OO$ denotes position and momentum of the first particle, $(\rr_1, \pp_1)$, and $\ii$ the analogically for the i-th particle. Delta distribution $\delta(\aaa-\ii)$ stands for $\delta(\rr_a-\rr_i)\delta(\pp^a-\pp^i)$. Derivative of $f$ with respect to $\fN$ is
\begin{equation}
 \frac{\partial f(\aaa)}{\partial \fN(\OO,\dots,\NN)} = \sum_{i=1}^N \delta(\aaa-\ii)
\end{equation}

Consider now two arbitrary functionals of the one-particle distribution function $f$, $A(f)$ and $B(f)$. Let us now evaluate the Liouville-Poisson bracket, \eqref{PB.eq.PB.Liouville}, of these two functionals.
\begin{eqnarray}
 \{A,B\}^{(L)} = \intd{\OO}\dots\intd{\NN}\intd{\aaa}\intd{\bb} \fN\frac{\partial A}{\partial f(\aaa)} \PB{\frac{f(\aaa)}{\partial \fN}}{\frac{\partial f(\bb)}{\partial \fN}}{CM}\frac{\partial B}{\partial f(\bb)}
\end{eqnarray}
Now using the definition \eqref{PB.eq.PB.CM}, this last equation becomes
\begin{equation}\label{PB.eq.B}
 \PB{A}{B}{B}= \intd{\rr}\intd{\pp}f(\rr,\pp)\left(\frac{\partial A_f}{\partial r^\alpha}\frac{\partial B_f}{\partial p_\alpha}-\frac{\partial B_f}{\partial r^\alpha}\frac{\partial A_f}{\partial p_\alpha}\right)
\end{equation}
where $A_f$ stands for $\frac{\partial A}{\partial f(\rr,\pp)}$. See \cite{dissertation} for detailed calculation. This last equation is exactly the Poisson bracket that generates the reversible part of the Boltzmann equation, referred to as the Boltzmann-Poisson bracket, see e.g. \cite{Grmela1997}. We have thus obtained that
\begin{equation}\label{PB.eq.LB}
 \PB{A}{B}{L} = \PB{A}{B}{B}
\end{equation}
for functionals $A$, $B$ dependent only\footnote{The functionals, of course, also depend on $\fN$ but only through the function $f$.} on the one-particle distribution function $f$.

In summary, starting with functionals dependent only on the one-particle distribution function, $f$, the Liouville Poisson bracket becomes the Boltzmann Poisson bracket. In other words, the reversible evolution governed by the Boltzmann equation is just a ``projection'' of the reversible evolution governed by the Liouville equation. Let us make this statement more rigorous.

\subsection{Geometric interpretation}\label{PB.sec.geoint}
Each Poisson bracket can be interpreted in terms of the corresponding Poisson bivector field as
\begin{equation}\label{eq.L}
 \{A,B\} = \left(L^{ij}\frac{\partial}{\partial x^i}\frac{\partial}{\partial x^j}\right)\left(d A, d B\right) = L^{ij}\frac{\partial A}{\partial x^i}\frac{\partial B}{\partial x^j},
\end{equation}
see e.g. \cite{Fecko}. Note that the indexes can be discrete or continuous (in which case the summation becomes integration) and that the derivatives are to be considered as functional derivatives. What is the relation between the Poisson bivector on the Liouville level of description and the Poisson bivector on the Boltzmann level?

The Boltzmann-Poisson bracket can be expressed in terms of its Poisson bivector as follows
\begin{equation}
 \PB{A}{B}{B} = \intd{\aaa}\intd{\bb} \frac{\partial A}{\partial f(\aaa)}L^{\aaa,\bb}_B\frac{\partial B}{\partial f(\bb)}
 \label{PB-B}
\end{equation}
and analogically the Poisson bivector generating the Liouville Poisson bracket is defined as
\begin{eqnarray}
 \PB{A}{B}{L} &=& \intd{\OO}\dots\intd{\NN}\intd{\OO'}\dots\intd{\NN'}\intd{\aaa}\intd{\bb}\nonumber\\
 &&\qquad\qquad\frac{\partial A}{\partial f(\aaa)}\frac{\partial f(\aaa)}{\partial \fN}L^{\fN,\fN'}_L\frac{\partial B}{\partial f(\bb)}\frac{\partial f(\bb)}{\partial \fN'}
\end{eqnarray}
where $\fN' = \fN(\OO',\dots,\NN')$ and $L^{\fN,\fN'}_L$ is the component of the Liouville Poisson bivector that provides coupling between $\OO,\dots,\NN$ and $\OO',\dots,\NN'$. Now from Eq. \eqref{PB.eq.LB} it follows that
\begin{equation}\label{PB.eq.LBLL}
 L^{\aaa,\bb}_B = \intd{\OO}\dots\intd{\NN}\intd{\OO'}\dots\intd{\NN'} \frac{\partial f(\aaa)}{\partial \fN}L^{\fN,\fN'}_L\frac{\partial f(\bb)}{\partial \fN'},
\end{equation}
which is the relation between the Poisson bivectors on the Liouville and Boltzmann levels.

Let us now have a look at relation \eqref{PB.eq.LBLL} from the geometrical point of view. Consider the manifold of state variables on a higher level of description (e.g. Liouville), and denote the state variables by $x^i$. On a lower level of description the state variables (denoted by $y^a$) are obtained by a projection
\begin{equation}
 \pi:\xx\rightarrow \yy(\xx).
\end{equation}
Poisson bivectors are twice contravariant antisymmetric tensors, and thus the projection transforms them formally as follows, see e.g. \cite{Fecko},
\begin{equation}\label{PB.eq.Lproj}
 \dL^{ab}\big|_{\pi(\xx)} = \frac{\partial \pi^a}{\partial x^i}\Big|_x  \uL^{ij}\big|_x\frac{\partial \pi^b}{\partial x^j}\Big|_x = \PB{\pi^a}{\pi^b}{higher},
\end{equation}
where $\dLL$ and $\uLL$ are the lower- and higher-level Poisson bivectors, respectively.
Formula \eqref{PB.eq.LBLL} is a particular realization of this last equation. The projection, however, does not need to be well defined, which is the reason for the word ``formally'', commented on later on.

In summary, obtaining the Poisson bracket on a lower level of description by evaluating the Poisson bracket on a higher level on functionals dependent only on state variables of the lower level is equivalent to projecting the Poisson bivector from the higher level of description to the lower level. Moreover, components of the Poisson bivector on the lower level of description are given by Poisson bracket on the higher level applied on components of the projector.

The Liouville-Poisson bracket generates reversible evolution in the sense of time-reversal transformation, see \cite{PRE2014}. Are the brackets derived from a reversible Poisson bracket also reversible? Consider state variables with definite parities with respect to the time-reversal transformation (TRT), which simply inverts velocities of all particles, see \cite{PRE2014}, i.e.
\begin{eqnarray}
 \TP{x^i} &=&  1 \mbox{ for } \IT(x^i) = x^i \mbox{, i.e. $x^i$ even, }\\
 \TP{x^i} &=&  -1 \mbox{ for } \IT(x^i) = -x^i \mbox{, i.e. $x^i$ odd, }
\end{eqnarray}
where $\TP{x^i}$ and $\IT(x^i)=\TP{x^i} x^i$ denote parity of variable $x^i$ and the action of TRT on the variable, respectively. Note that TRT satisfies $\IT\circ\pi=\pi\circ\IT$ as this is in fact the definition of TRT action on the more macroscopic level (and hence we do not distinguish the two distinct TRT mappings on the lower and higher level of description). Additionally, the commutation relation yields
\begin{equation*}
	\frac{\partial \IT^a}{\partial y^b}\Big|_{\pi(x)}	\frac{\partial \pi^b}{\partial x^i}\Big|_{\xx}=	\frac{\partial \pi^a}{\partial x^j}\Big|_{\IT(x)}	 \frac{\partial \IT^j}{\partial x^i}\Big|_{\xx}
\end{equation*}
and as $\frac{\partial \IT^j}{\partial x^i}\Big|_{\xx}=\delta_i^j \TP{x^i}$ and $\frac{\partial \IT^a}{\partial y^b}\Big|_{\xx}=\delta_b^a \TP{y^b}$, we have
\begin{equation} \label{aux.1}
	\frac{\partial \pi^a}{\partial x^i}\Big|_{\IT(\xx)}=\TP{x^i}\TP{y^a}\frac{\partial \pi^a}{\partial x^i}\Big|_{\xx}.
\end{equation}

Suppose, moreover, that the Poisson bracket on the higher level of description generates reversible evolution (as for example the Liouville Poisson bracket does), which means that
\begin{equation}\label{eq.Lrev}
 \uL^{ij}|_{\IT(\xx)} = -\TP{x^i}\TP{x^j} \uL^{ij}|_{\xx},
\end{equation}
as shown in \cite{PRE2014}. The Poisson bivector at the lower level evaluated at inverted coordinates (i.e. after application of TRT) becomes
\begin{eqnarray}
 \dL^{ab}|_{\IT(\yy)} &=& \dL^{ab}|_{\IT\circ\pi(\xx)}=\dL^{ab}|_{\pi\circ\IT(\xx)}  \nonumber\\
 &=&\frac{\partial \pi^a}{\partial x^i}\Big|_{\IT(\xx)}\uL^{ij}\big|_{\IT(\xx)}\frac{\partial \pi^b}{x^j}\Big|_{\IT(\xx)} \nonumber\\
 &\stackrel{\eqref{aux.1}}{=}&\TP{y^a}\TP{y^b}\TP{x^i}\TP{x^j}\frac{\partial \pi^a}{\partial x^i}\Big|_{\xx}\uL^{ij}\big|_{\IT(\xx)}\frac{\partial \pi^b}{x^j}\Big|_{\xx}\nonumber\\
 &\stackrel{\eqref{eq.Lrev}}{=}&-\TP{y^a}\TP{y^b}\frac{\partial \pi^a}{\partial x^i}\Big|_{\xx}\uL^{ij}\big|_{\xx}\frac{\partial \pi^b}{x^j}\Big|_{\xx}\nonumber\\
 &=&-\TP{y^a}\TP{y^b}\dL^{ab},
\end{eqnarray}
which means that any Poisson bivector obtained by projection of a Poisson bivector generating reversible evolution generates only reversible evolution as well. Since the Poisson bivector of classical mechanics or Liouville equation generates reversible evolution, all the Poisson bivectors (or Poisson brackets) derived by the projection from the Liouville Poisson bivector generate only reversible evolution.

Let us now return to the word ``formally'' used when stating that the Poisson bracket on the lower level is obtained by a projection. The projector $\pi$ is no diffeomorphism, and thus the push-forward mapping bivectors on $M$ to bivectors on $N$ does not exist in general, see \cite{Fecko}, and without providing any additional knowledge about the system under consideration there is no a priori guarantee that the result of the projection is well defined. An example of the additional knowledge that ensures possibility of the projection is a symmetry permitting a symplectic or Poisson reduction, see \cite{Fecko} or \cite{Abraham-Marsden}. If, on the other hand, the result of the projection given by Eq. \eqref{PB.eq.Lproj} only depends on state variables of the lower level, $y$, the projection is well defined and one may use it as the Poisson bivector on the lower level. The explicit calculation of the Poisson bivector (or Poisson bracket) on the lower level of description has to be carried out anyway and so the checking whether it only depends on state variables of the lower level does not bring any additional computational effort.

But is the Jacobi identity guaranteed for the result of the projection? Firstly, consider smooth functions on $M$, $\FFF(M)$, which are also referred to as the algebra of observables on $M$. Functions from $\FFF(M)$ indeed form an (associative) algebra, since they form a linear space and the standard pointwise multiplication is the bilinear operation necessary for $\FFF(M)$ to be an algebra. Moreover, Poisson bracket of two functions from $\FFF(M)$ stays in $\FFF(M)$ and so $\FFF(M)$ is even a Lie algebra, referred to as an algebra of observables, see \cite{Fecko}.

Secondly, consider functions on $M$ that are projectable onto $N$,
\begin{equation}
 \FFF_\pi(M) = \{ A\in \FFF(M): \pi(x_1) = \pi(x_2) \Rightarrow A(x_1) = A(x_2)\}
\end{equation}
Functions $\FFF_\pi(M)$ form an algebra isomorphic to the algebra of functions on $N$, $\FFF(N)$. Indeed, there is a bijective mapping between $\FFF_\pi(M)$ and $\FFF(N)$ which respects the bilinear operation (pointwise multiplication) given by $A^H \sim A^L \circ \pi$, where $A^H \in \FFF_\pi(M)$  and $A^L \in \FFF(N)$.

So far we have shown that $\FFF_\pi(M)$ is an algebra isomorphic with $\FFF(N)$. Let us now introduce a Poisson structure inherited from the algebra of observables $\FFF(M)$ so that $\FFF_\pi(M)$ also becomes an algebra of observables. Consider two functions $A^L$ and $B^L$ from $\FFF_\pi(M)$ and define their Poisson bracket as
\begin{equation}\label{PB.eq.ABL}
 \{A^L,B^L\}^L \stackrel{def}{=} \{A^L,B^L\}^H =\uL^{ij} \frac{\partial A^L}{\partial x^i}\frac{\partial B^L}{\partial x^j}
 =\underbrace{\uL^{ij} \frac{\partial \pi^a}{\partial x^i}\frac{\partial \pi^b}{\partial x^j}}_{\dL^{ab}}\frac{\partial A^L}{\partial y^a}\frac{\partial B^L}{\partial y^b} = \PB{A}{B}{L}
\end{equation}
Does this expression give a function from $\FFF_\pi(M)$? Not in general since one has to check that the Poisson bivector on the lower level of description, $\dL^{ab}$, does not depend on $x$ but only on $y$. Note, however, that such calculation has to be carried out anyway when one wishes to express the tensor explicitly. But when the resulting bivector field only depends on state variables $y$, the expression $\{A^L,B^L\}^L$ indeed is in $\FFF_\pi(M)$, and due to the properties of the Poisson bracket $\{\bullet,\bullet\}^H$ (antisymmetry, Jacobi identity) the $\{\bullet,\bullet\}^L$ is indeed a Poisson bracket. Since $\FFF_\pi(M)$ and $\FFF(N)$ are isomorphic, the Poisson bracket works on $\FFF(N)$ as well, and we have obtained an algebra of observables $\FFF(N)$ with the Poisson bracket $\{\bullet,\bullet\}^L$ inherited from $\FFF(M)$. The expression for the Poisson bracket on the lower level of description and the expression for the new Poisson bivector are exactly the same as in the illustration demonstrating projection from the Liouville level to the Boltzmann level as well as in all other illustrations of the procedure in this paper, and so one does not need to check whether the derived brackets really are Poisson brackets (fulfilling antisymmetry and Jacobi identity) because that is satisfied automatically when the projection can proceed (the resulting Poisson bivector depends only on the $y$ variables).

In summary, it has been shown that when starting with a Poisson bracket generating reversible evolution on a higher level of description, there is a natural way to construct a Poisson bracket on a lower level of description. The new Poisson bracket indeed satisfies all the properties of Poisson brackets, and it generates only reversible evolution. The passage to the lower level is given by defining a projection from state variables on the higher level to state variables on the lower level. During the passage one has to check that the constructed Poisson bivector  depends only on the state variables of the lower level. If this is true, the passage is complete and the dynamics on the lower level  is derived from the reversible dynamics on the higher level, it is reversible, and the Poisson bracket indeed fulfills all the necessary properties (antisymmetry and Jacobi identity).

Note that reversible
 evolution on a lower level of description is often derived from the Liouville equation by means of statistical physics, see \cite{Ottinger}. For instance, Turkington \cite{Turkington} recently developed  an optimization procedure leading from Liouville equation to GENERIC, and the Poisson bracket is given by Eq. \eqref{PB.eq.Lproj}. Therefore, the way Poisson brackets are constructed in this manuscript is compatible with the way they are constructed from Liouville equation in statistical physics.

\subsection{Particular examples of the hierarchy}

In the following subsections we shall provide some particular illustrations of the above geometric approach. Consequently, all the listed levels of descriptions can be regarded as tightly linked one to another and all consistently derived from the first principles. Later in the text we also show examples that are outside of this unification theory and propose a generalization for certain situations.

\subsubsection{From Boltzmann to classical hydrodynamics}\label{sec.PB.CH}
Having derived the Boltzmann-Poisson bracket in Sec. \ref{PB.sec.B}, we are able to derive the hydrodynamic Poisson bracket in a convenient way. One could, of course, start with the Liouville Poisson bracket and go directly to hydrodynamics, but the calculation would be longer than when starting from the Boltzmann Poisson bracket. Let us, therefore, introduce the projection from the Boltzmann level of description to the level of classical hydrodynamics.

The hydrodynamic state variables are density, momentum density and entropy density
\begin{subequations}
\begin{eqnarray}
 \rho(\rr_a)&=&\intd\rr\intd\pp m f(\rr,\pp)\delta(\rr-\rr_a) \label{PB.eq.CH.rho}\\
 u_i(\rr_a)&=&\intd\rr\intd\pp p_i f(\rr,\pp)\delta(\rr-\rr_a)\label{PB.eq.CH.u}\\
 \label{PB.eq.CH.s}s(\rr_a)&=&\intd\rr\intd\pp \sff(f(\rr,\pp))\delta(\rr-\rr_a)
\end{eqnarray}
\end{subequations}
where $m$ is mass of one particle and entropy density $\sff$ is a positive smooth real-valued function of the distribution function. Therefore, derivatives of the hydrodynamic state variables with respect to $f(\rr,\pp)$ are
\begin{subequations}\label{PB.eq.CHder}
\begin{eqnarray}
 \frac{\partial \rho(\rr_a)}{\partial f(\rr,\pp)} &=& m \delta(\rr-\rr_a),\\
 \label{PB.eq.CHder.u}\frac{\partial u_i(\rr_a)}{\partial f(\rr,\pp)} &=& p_i \delta(\rr-\rr_a)\mbox{ and}\\
 \frac{\partial s(\rr_a)}{\partial f(\rr,\pp)} &=& \sff'(f(\rr,\pp))\delta(\rr-\rr_a),
\end{eqnarray}
\end{subequations}
\begin{equation}\label{eq.CH.Ax}
 \frac{\partial A}{\partial f(\rr,\pp)} = m A_\rho + p_i A_{u_i} + \sigma' A_s.
\end{equation}
Let us now construct the hydrodynamic Poisson bracket. Plugging Eq. \eqref{eq.CH.Ax} into the Boltzmann-Poisson bracket, \eqref{PB.eq.B}, we obtain
\begin{subequations}\label{B-CH}
\begin{eqnarray}
 \PB{A}{B}{B} &=& \intd\rr\intd\pp f \left(\frac{\partial}{\partial r^k}\left(m A_\rho + p_i A_{u_i} + \sigma' A_s\right)\frac{\partial}{\partial p_k}\left(m B_\rho + p_j B_{u_j} + \sigma' B_s\right)-\dots\right)\nonumber\\
 \label{B-CH1}&=&\intd\rr\intd\pp f\left(\left(m\frac{\partial A_\rho}{\partial r^k} + p_i \frac{\partial A_{u_i}}{\partial r^k}\right)B_{u_k}-\dots\right)\\
 \label{B-CH2}&&+\intd\rr\intd\pp f \left(\frac{\partial \sigma' A_s}{\partial r^k} B_{u_k}-\dots\right)\\
 \label{B-CH3}&&+\intd\rr\intd\pp f\left(\left(m \frac{\partial A_\rho}{\partial r^k} + p_i \frac{\partial A_{u_i}}{\partial r^k}\right)\frac{\partial}{\partial p_k}\left(\sigma' B_s\right)-\dots\right)\\
 \label{B-CH4}&&+\intd\rr\intd\pp f \left(\frac{\partial}{\partial r^k}\left(A_s \sigma'\right)\frac{\partial}{\partial p_k}\left( B_s \sigma'\right)-\dots\right).
\end{eqnarray}
\end{subequations}
The dots stand for the antisymmetric complement, where $A$ and $B$ are swapped.
Term \eqref{B-CH1} can be rewritten as
\begin{subequations}
\begin{equation}
\intd\rr \rho(\partial_k A_\rho B_{u_k}- \partial_k B_\rho A_{u_k}) + u_i(\partial_k A_{u_i} B_{u_k}- \partial_k B_{u_i} A_{u_k}),
\end{equation}
which constitutes the first two terms of the final hydrodynamic Poisson bracket.
Term \eqref{B-CH2} becomes
\begin{multline}\label{B-CH22}
 -\intd\rr\intd\pp \underbrace{\frac{\partial f}{\partial r^k}\sigma'}_{=\frac{\partial \sigma}{\partial r^k}} A_s B_{u_k}+\dots -\intd\rr\intd\pp f\sigma' A_s \frac{\partial B_{u_k}}{\partial r^k}+\dots\\
=  \intd\rr s (\partial_k A_s B_{u_k}-\partial_k B_s A_{u_k}) + \intd\rr\intd\pp \sigma A_s \frac{\partial B_{u_k}}{\partial r^k}-\dots \\
-\intd\rr\intd\pp f\sigma' A_s \frac{\partial B_{u_k}}{\partial r^k}+\dots,
\end{multline}
where the first term following the equality sign is the last term of the hydrodynamic Poisson bracket. All the other terms will cancel with the remaining terms of \eqref{B-CH}. Term \eqref{B-CH3} becomes
\begin{multline}
 -\intd\rr\intd\pp \underbrace{\frac{\partial f}{\partial p_k}\sigma'}_{\frac{\partial \sigma}{\partial p_k}} B_s \left(m\frac{\partial A_\rho}{\partial r^k} + p_i \frac{\partial A_{u_i}}{\partial r^k}\right)+\dots - \intd\rr\intd\pp f\frac{\partial A_{u_k}}{\partial r^k}\sigma' B_s + \dots\\
 = \intd\rr\intd\pp \sigma B_s \frac{\partial A_{u_k}}{\partial r^k}-\dots - \intd\rr\intd\pp f\frac{\partial A_{u_k}}{\partial r^k}\sigma' B_s + \dots,
\end{multline}
which cancels with the second and third term following the equality sign in \eqref{B-CH22}. Finally, term \eqref{B-CH4} can be rewritten as
\begin{multline}
 \intd\rr\intd\pp f \left(\frac{\partial A_s \sigma'}{\partial r^k} B_s \frac{\partial \sigma'}{\partial p_k}-\dots\right) = \intd\rr\intd\pp f \frac{\partial A_s}{\partial r^k}B_s \underbrace{\sigma'\frac{\partial \sigma'}{\partial p_k}}_{=\frac{\partial \Sigma'}{\partial p_k} \mbox{ for } \Sigma' = \frac{1}{2} (\sigma')^2}-\dots\\
=\intd\rr\intd\pp -\frac{\partial f}{\partial p_k} \frac{\partial A_s}{\partial r^k}B_s \Sigma' + \dots \\
=-\intd\rr\intd\pp \frac{\partial \Sigma}{\partial p_k} \left(\frac{\partial A_s}{\partial r^k}B_s -\frac{\partial B_s}{\partial r^k}A_s\right) = 0,
\end{multline}
by integration by parts.
\end{subequations}

In summary, all the terms that remain from bracket \eqref{B-CH} form the Poisson bracket of classical hydrodynamics,
\begin{eqnarray}\label{HB}
 \PB{A}{B}{CH} &=& \intd\rr \rho\left(\partial_i A_\rho B_{u_i}-\partial_i B_\rho A_{u_i}\right)+\nonumber\\
 &&+\intd\rr u_i\left(\partial_j A_{u_i} B_{u_j}-\partial_j B_{u_i} A_{u_j}\right)+\nonumber\\
 &&+\intd\rr s\left(\partial_i A_s B_{u_i}-\partial_i B_s A_{u_i}\right),
\end{eqnarray}
 see also \cite{Marsden-Ratiu-Weinstein}, \cite{Ottinger} or  Appendix of \cite{MG2014}, where the calculations needed to pass  from the Boltzmann Poisson bracket to the hydrodynamic bracket \eqref{HB} are made somewhat differently.


\subsubsection{From Liouville to binary Boltzmann}
Consider now that there are particles of two different species in the isolated system. Their positions and momenta are denoted by $\OO, \dots, \NN$ and $\tOO, \dots, \tNN$. The overall $(N+\tilde{N})$-particle distribution function, which evolves according to Liouville equation, is denoted by $\fNN$. One-particle distribution functions of the two species are then defined by the following projections.
\begin{subequations}\label{eq.LBproj}
\begin{eqnarray}
 f(\aaa) &=& \intd{\OO}\dots\intd{\NN}\fNN\sum_{i=1}^N\delta(\ii-\aaa)\Rightarrow \frac{\partial f(\aaa)}{\partial \fNN} = \sum_{i=1}^N \delta(\ii-\aaa)\\
 \tf(\aaa) &=& \intd{\OO}\dots\intd{\NN}\fNN\sum_{\ti=1}^N\delta(\tii-\aaa)\Rightarrow \frac{\partial \tf(\aaa)}{\partial \fNN} = \sum_{\ti=1}^N\delta(\tii-\aaa).
\end{eqnarray}
\end{subequations}
Poisson bracket governing evolution of these one-particle distribution functions is the Poisson bracket governing evolution of binary mixtures within the context of kinetic theory. Let us now calculate the bracket.

Introducing also $N$-particle distribution functions of the two species,
\begin{subequations}
\begin{eqnarray}
 \fN(\OO,\dots,\NN) &=& \intd{\tOO}\dots\intd{\tNN}\fNN\mbox{ and}\\
 \tfN(\tOO,\dots,\tNN) &=& \intd{\OO}\dots\intd{\NN}\fNN,
\end{eqnarray}
\end{subequations}
Poisson bracket \eqref{PB.eq.PB.Liouville} can be rewritten as (see \cite{dissertation} for details)
\begin{equation}
 \PB{A}{B}{L} = \PB{A}{B}{L(N)} + \PB{A}{B}{L(\tilde{N})}
\end{equation}
where $\PB{A}{B}{L(N)}$ is the Liouville Poisson bracket of the two functionals constructed from the $N$-particle distribution function $\fN$, and $\PB{A}{B}{L(\tilde{N})}$ the Liouville Poisson bracket constructed from $\tfN$. But as we have already seen in Sec. \ref{PB.sec.B}, these two Poisson brackets are equal to the Boltzmann Poisson brackets constructed from $\fN$ and $\tfN$, respectively. Consequently, the Poisson bracket governing reversible evolution of two one-particle distribution functions is
\begin{equation}\label{PB.eq.PBBB}
 \PB{A}{B}{B\times \tilde{B}} = \PB{A}{B}{B} + \PB{A}{B}{\tilde{B}}
\end{equation}
where the latter bracket is the Boltzmann-Poisson bracket with $\tf$ instead of $f$.


\subsubsection{From binary Boltzmann to binary hydrodynamics}
Poisson bracket of binary hydrodynamics, where the each of the fluids is described by its density, momentum density and entropy density, can be obtained from the binary Boltzmann-Poisson bracket straightforwardly. The projection is as follows:
\begin{subequations}\label{PB.eq.BH.state}
\begin{eqnarray}
 \rho(\ra) &=& \intd{\rr}\intd{\pp}f(\rr,\pp)m\delta(\rr_a-\rr),\\
 u_i(\ra) &=& \intd{\rr}\intd{\pp}f(\rr,\pp)p_i\delta(\rr_a-\rr),\\
 s(\ra) &=& \intd{\rr}\intd{\pp} \sff(f(\rr,\pp))\delta(\rr_a-\rr),\\
 \label{PB.eq.trho}\trho(\ra) &=& \intd{\trr}\intd{\tpp}\tf(\trr,\pp)m\delta(\rr_a-\trr),\\
 \tu_i(\ra) &=& \intd{\trr}\intd{\tpp}\tf(\trr,\tpp)\tp_i\delta(\rr_a-\trr)\mbox{ and}\\
 \label{PB.eq.ts}\ts(\ra) &=& \intd{\trr}\intd{\tpp}\tsf(\tf(\trr,\tpp))\delta(\rr_a-\trr).
\end{eqnarray}
\end{subequations}

Consider now two functionals dependent only on these state variables. Poisson bracket \eqref{PB.eq.PBBB} does not provide coupling between the two different species. Therefore, the bracket applied to the two functionals consists of two parts, each of which is obtained the same way as the hydrodynamic Poisson bracket was obtained from Boltzmann Poisson bracket. Therefore, the Poisson bracket describing evolution of state variables \eqref{PB.eq.BH.state} is
\begin{equation}\label{PB.eq.PBCHCH}
 \PB{A}{B}{CH\times \widetilde{CH}} = \PB{A}{B}{CH} + \PB{A}{B}{\widetilde{CH}}
\end{equation}
where the latter Poisson bracket is the standard hydrodynamic Poisson bracket but with state variables \eqref{PB.eq.trho}-\eqref{PB.eq.ts}.

In summary, a binary mixture with state variables \eqref{PB.eq.BH.state} evolves according to Poisson bracket \eqref{PB.eq.PBCHCH}, which is simply the sum of hydrodynamic Poisson brackets, one for each species. Note that this description of binary mixtures typically represents mixtures with different temperatures (as in cold plasma, where electrons have different temperature than ions, see \cite{Chen}). Indeed, as there are two entropies, there are also two temperatures (derivatives of the entropies with respect to energy).

\subsubsection{From binary hydrodynamics to CIT}\label{sec.PB.CIT}
Let us approach an even lower level of description, where only densities, total momentum, and total entropy density constitute the state variables. The projection is then given by
\begin{subequations}
\begin{eqnarray}
 \rho(\rr) &=& \rho(\rr),\\
 \trho(\rr) &=& \trho(\rr),\\
 \uT_i(\rr) &=& u_i(\rr)+\tu_i(\rr)\mbox{ and}\\
 \sT(\rr) &=& s(\rr) + \ts(\rr).
\end{eqnarray}
\end{subequations}
This level of description is quite important since it is the level where Classical Irreversible Thermodynamics (CIT), see \cite{dGM} or \cite{Lebon-Understanding}, takes place. Indeed, the state variables are the same as within CIT. Note that the CIT Poisson bracket was already introduced in \cite{Ottinger2009} using Dirac constraints. The following derivation can be regarded as a shortcut to the Poisson bracket.

Derivatives of a functional dependent only on $(\rho,\trho,\uuT,\sT)$ are then
\begin{equation}
 A_\rho = A_\rho, A_{\trho} = A_{\trho}, A_{\uu} = A_{\tu} = A_{\uuT} \mbox{ and } A_s = A_{\ts} = A_s.
\end{equation}
Plugging these relations into bracket \eqref{PB.eq.PBCHCH} yields the CIT Poisson bracket,
\begin{eqnarray}
 \PB{A}{B}{CIT} &=& \intd\rr \rho(\partial_i A_{\rho} B_{\uT_i}-\partial_i B_{\rho} A_{\uT_i})\nonumber\\
 &&+ \intd\rr \trho(\partial_i A_{\trho} B_{\uT_i}-\partial_i B_{\trho} A_{\uT_i})\nonumber\\
 &&+\intd\rr \uT_i (\partial_j A_{\uT_i} B_{\uT_j}-\partial_j B_{\uT_i} A_{\uT_j})\nonumber\\
 &&+\intd\rr \sT (\partial_j A_{\sT} B_{\uT_j}-\partial_j B_{\sT} A_{\uT_j}).
\end{eqnarray}

Taking energy equal to
\begin{equation}
 E = \int d\rr \frac{(\uuT)^2(\rr)}{2(\rho(\rr)+\trho(\rr))} + \eps(\rho(\rr),\trho(\rr), \sT(\rr)),
\end{equation}
where internal energy density $\eps$ is just a real-valued function of the state variables, the evolution equations become
\begin{subequations}
\begin{eqnarray}
 \frac{\partial \rho}{\partial t}&=& -\partial_i\left(\rho \frac{\uT_i}{\rho+\trho}\right) \\
 \frac{\partial \trho}{\partial t}&=& -\partial_i\left(\trho \frac{\uT_i}{\rho+\trho}\right) \\
 \frac{\partial \uT_i}{\partial t}&=&-\partial_j\left(\frac{\uT_i\uT_j}{\rho+\trho}\right)-\partial_i p\\
 \frac{\partial \sT}{\partial t} &=& -\partial_i\left(\sT\frac{\uT_i}{\rho+\trho}\right)
\end{eqnarray}
\end{subequations}
where pressure is expressed in terms of state variables through
\begin{equation}
 p = -\eps+\rho\frac{\partial \eps}{\partial \rho} + \trho\frac{\partial \eps}{\partial \trho} + \sT\frac{\partial \eps}{\partial \sT},
\end{equation}
which is clearly equivalent to the reversible part of CIT evolution equation, see e.g. \cite{dGM}. In other words, it has been shown how the reversible part of evolution within CIT can be derived by projection from the Liouville equation.

One could also continue in the reduction to a lower level, where only the total density
\begin{equation}
 \rho^T = \rho + \trho
\end{equation}
total momentum $\uu^T$ and total entropy $\sT$ are present. This is an alternative (and longer) way to obtain the classical hydrodynamic Poisson bracket.

Furthermore, one could project the total momentum to a constant. This is the passage to the level of mechanical equilibrium introduced for example in \cite{dGM}. The Poisson bracket, however, completely disappears during the passage, since the CIT Poisson bivector does not provide coupling between densities and total entropy, which are the only state variables that remain on the level of mechanical equilibrium.

\subsubsection{From non-symmetric Liouville to symmetric Liouville}\label{L-LS}
Hitherto, we have not required the N-particle distribution functions to be symmetric with respect to permutations of particles. Since, however, elementary particles are indistinguishable, let us pass to the distribution functions symmetric with respect to permutations of particles. The symmetric distribution function is given by
\begin{multline}\label{eq.fNS}
 \fNS (\OO,\dots,\NN) = \\
 =\intd\OO'\dots\intd\NN \frac{1}{N!}\sum_P \left(\delta(\OO-P(\OO')\dots\delta(\NN-P(\NN'))\right)\fN(\OO',\dots,\NN'),
\end{multline}
where $P$ denotes permutations of labels $1,\dots,N$.

The Liouville-Poisson bracket, Eq. \eqref{PB.eq.PB.Liouville}, then becomes (details in ES)
\begin{multline}\label{eq.LSym.PB.final}
 \PB{A(\fNS)}{B(\fNS)}{LS} = \\
 =\intd\OO\dots\intd\NN\fNS(\OO,\dots,\NN)\PB{\frac{\partial A}{\partial \fNS(\OO,\dots,\NN)}}{\frac{\partial B}{\partial \fNS(\OO,\dots,\NN)}}{CM},
\end{multline}
which is the Poisson bracket for symmetric $n$-particle distribution functions. The Poisson bivector indeed only depends on $\fNS$, and thus the projection to symmetric distributions defines a Lie algebra.

The above procedure can be also seen in a more geometric way by regarding relations between the Lie algebras of n-and-one-particle distribution functions as shown in ES.

Regarding also irreversible evolution, which is not described by the Poisson bracket, one could seek for a dissipation potential such that loss of particle identity is revealed as an irreversible process. Such evolution would then lead from the non-symmetric N-particle distribution function to the symmetric counterpart.

In summary, symmetrization provides passage between the level of general N-particle distribution functions, governed by Liouville-Poisson bracket, and the level of symmetric N-particle distributions, governed by the Liouville Poisson-bracket with $\fNS$ instead of $\fN$. Let us refer to the level of description where only symmetric N-particle distributions constitute the state variables as to the symmetric Liouville level (LS).

\section{Projections incompatible with the hierarchy}\label{sec.out}

\subsection{From binary hydrodynamics to EIT}\label{sec.PB.EIT}
Let us analyze a level of description between binary hydrodynamics and CIT, where only one (the total) entropy  density is defined. This situation corresponds to an extended irreversible thermodynamic (EIT) level within mixture theory as discussed in \cite{Pavelka2014}. The projection is given by
\begin{subequations}
\begin{eqnarray}\label{eq.EIT.state}
 \rho(\rr) &=& \rho(\rr),\\
 \trho(\rr) &=& \trho(\rr),\\
 u_i(\rr) &=& u_i(\rr),\\
 \tu_i(\rr)&=&\tu_i(\rr)
\end{eqnarray}
\end{subequations}
and a total entropy density relation of which to binary hydrodynamic variables $\sT=\sT(\rho,\trho,\uu,\tuu)=s+\ts+K(\rho,\trho,\uu,\tuu)$ is to be identified. Motivation for the unknown function $K$ lies in the work \cite{Pavelka2014}, where it has been identified as minus the kinetic energy of diffusion over temperature. Let us denote the vector of binary hydrodynamics variables as $BH$ and the EIT variables as $EIT$.

The detailed calculations, that can be found in ES, lead to the conclusion that the discussed EIT theory of mixtures is out of the hierarchy of Poisson brackets. Although it does not mean that the theory is incompatible with GENERIC, as there still might be a Poisson bracket governing reversible evolution of the theory, it seems to be rather problematic to construct such a Poisson bracket. The difficult task is to couple two momenta and with only one entropy which thus has to be convected by the two momenta, most likely leading to violation of Jacobi identity. On the other hand, even if it turns out that there is no Poisson bracket expressing reversible evolution in the theory of mixtures, the theory may be regarded as an useful approximation of the two-entropy (two-temperature) setting. Indeed, it was shown in paper \cite{Pavelka2014} that the entropy-less part of the Poisson bracket of binary hydrodynamics generates reversible evolution in the theory of mixtures in the isothermal case.

\subsection{From Liouville to two-point BBGKY}
\label{Lto2BBGKY}
The symmetric Liouville level can be rewritten in terms of the whole BBGKY hierarchy \cite{Bogoliubov1, Bogoliubov2, Kirkwood1, Kirkwood2, Born-Green}, and the hierarchy thus acquires Hamiltonian structure as shown in \cite{Marsden-BBGKY}. However, it is important to consider all the levels of the BBGKY hierarchy at once. For example the evolution equation of the two-particle level of BBGKY contains also a contribution from the three-particle distribution function, etc. It is shown in ES that the projection from symmetric Liouville distribution functions to only one-and-two-particle distribution functions does not yield a Poisson bracket. Only the projection from symmetric Liouville distribution functions to one-particle distribution functions, given by Eq. \eqref{PB.eq.ffN}, provides a Poisson bracket, see 
\ref{sec.LB.groups} for more details, where the geometric reason is identified.

Note that the projection forms a Lie algebra only in case of the Boltzmann equation, i.e. when projecting only to the one-particle distribution functions, since then no interactions between particles play a role. Indeed, even in the classical BBGKY hierarchy evolution equations of distribution functions depend on higher-order distribution functions via interactions between particles (interaction part of Hamiltonian), but when no interaction is present, the evolution equations become closed. This does not mean, however, that Boltzmann equation does not describe systems with interactions, see e.g. \cite{Miroslav-PhysicaD}.

According to the explicit calculation in ES, the projection from Liouville level to two-point BBGKY is not preserving the Poisson structure, in other words, the projection is not a Poisson map. It is possible to observe the obstruction on the projection being Poisson in a pure algebraic way, see Remark \ref{Remark.L.2BBGKY} in \ref{sec.GCE.groups}.

\section{Generalization of the hierarchy}
\subsection{Grand-canonical ensemble of Liouville equations}\label{sec.GCE}
The standard Liouville equation describes evolution of a fixed number of indistinguishable particles. It is possible to formulate analogical evolution when the number of particles is not fixed? A similar question is resolved in equilibrium statistical physics when passing from canonical ensemble to grand-canonical ensemble, see e.g. \cite{Gibbscw} or \cite{Kubo} and in the quantum field theory, e.g. \cite{Landau9}. The passage is interpreted by Gibbs as taking an ensemble of systems with different numbers of particles instead of only one.  Motivated by this observation, a grand-canonical ensemble (GCE) hierarchy has been introduced in \cite{Miroslav-EKT} and \cite{Miroslav-external}. The purpose of this section is to review and further develop the GCE hierarchy.

Consider isolated systems consisting of $1,2,3,\dots$ particles, which have all the same volume. Each of the systems evolves according to the one-, two-, three-, etc. particle symmetric Liouville equation, respectively, and the evolution is expressed by the corresponding symmetric grand-canonical distribution functions. For example, the average number of particles, average energy and average entropy can be expressed as
\begin{eqnarray}\label{eq.EN.f}
 N &=& \sum_{n=1}^\infty \frac{1}{n!}\intd\OO\dots\intd\nnn n \en(\OO,\dots,\nnn)\\
 E &=& \sum_{n=1}^\infty E_n =\sum_{n=1}^\infty \underbrace{\frac{1}{n!} \intd\OO\dots\intd\nnn H_n(\OO,\dots,\nn) \en(\OO,\dots,\nnn)}_{E_n}\\
 \label{eq.GCE.S}S &=& -k_B\sum_{n=1}^\infty \frac{1}{n!}\intd\OO\dots\intd\nnn \sigma_n(\en)
\end{eqnarray}
where $H_n$ is a $n-$particle Hamiltonian
\begin{equation}
 H_n(\OO,\dots,\nnn) = \sum_{i=1}^n h_1(\ii) + \frac{1}{2}\sum_{i=1}^n\sum_{\stackrel{j=1}{j\neq i}}^n h_2(\ii,\jj)+\dots,
\end{equation}
and $\sigma_n$ stands for a real-valued concave function of one variable.

In thermodynamic equilibrium the distribution functions are given by
\begin{equation}\label{en.eq}
 \en(\OO,\dots,\nnn) = \frac{1}{h^{3n}}e^{-\beta p V}\exp(\beta\mu n - \beta H_n(\OO,\dots,\nnn))
\end{equation}
where $\beta = 1/k_B T$, $T$ temperature, $h$ is Planck constant, $k_B$ Boltzmann constant, $p$ pressure, $V$ volume and $\mu$ chemical potential. The functions are normalized by
\begin{equation} \label{norm-e}
 1 = \sum_{n=0}^\infty \frac{1}{n!} \intd\OO\dots\intd\nnn \en(\OO,\dots,\nnn),
\end{equation}
see \cite{Cargese} or \cite{Kubo}. We suppose that the relation between the grand-canonical distribution functions and the symmetric Liouville distribution functions is
\begin{equation}\label{eq.enfn}
 \en = \en^0 \fns
\end{equation}
where the prefactor $\en^0$ expresses the probability that there is $n$ particles in the system.

Let us now transform the grand-canonical distribution functions as suggested by Melville Green in \cite{Cargese}. The transformation also appears naturally when regarding the Lie algebra structure of the grand-canonical ensemble as discussed in 
\ref{sec.GCE.groups}. The new distribution functions are
\begin{subequations}\label{eq.f.rho}
\begin{eqnarray}
 \rho_1(\OO) &=& \sum_{n=1}^\infty\frac{1}{(n-1)!}\intd\TT\dots\intd\nnn \en(\OO,\dots\nnn)\\
 \rho_2(\OO,\TT) &=& \sum_{n=2}^\infty\frac{1}{(n-2)!}\intd\TTT\dots\intd\nnn \en(\OO,\dots\nnn)
\end{eqnarray}
\end{subequations}
and analogically for higher $n$. The average number of particles and the average energy thus become
\begin{subequations}\label{eq.EN.rho}
\begin{eqnarray}
 N &=& \intd\OO \rho_1(\OO)\mbox{ and}\\
 \label{eq.E.rho}E &=& \intd\OO h_1(\OO)\rho_1(\OO) + \frac{1}{2}\intd\OO\intd\TT h_2(\OO,\TT)\rho_2(\OO,\TT)+\dots
\end{eqnarray}
\end{subequations}
Note that energy is now conveniently split into contribution from n-particle interactions instead of the n-particle Hamiltonians, which simplifies the formula to great extent. Such splitting can be regarded as a motivation for introducing the new distribution functions as can be seen for example in \ref{sec.GCE.groups}.

Transformation \eqref{eq.f.rho} can be inverted as
\begin{equation}\label{eq.rho.f}
 \ei(\OO,\dots,\ii) = \sum_{n=i}^\infty \frac{(-1)^{(n-i)}}{(n-i)!}\intd(\ii+\OO)\dots\intd\nnn \rho_n(\OO,\dots,\nnn),
\end{equation}
and that derivative with respect to $\en$ of a functional dependent on $\rho_1,\rho_2,\dots$ is
\begin{equation}\label{eq.GCE.Afn}
 \frac{\partial A}{\partial \en(\OO,\dots,\nnn)} = \frac{1}{n!}\sum_{i=1}^n \frac{\partial A}{\partial \rho_1(\ii)} + \frac{1}{n!} \sum_{i=1}^n \sum_{\stackrel{j=1}{j\neq i}}^n \frac{\partial A}{\rho_2(\ii,\jj)} + \dots
\end{equation}

What is the physical meaning of transformation \eqref{eq.f.rho}? For example $\rho_1(\OO)$ can be interpreted as the probability that there is a particle at position $\OO$ regardless of the overall number of particles in the system. Similarly, $\rho_2(\OO,\TT)$ can be interpreted as the probability that there are two particles at those two positions regardless the number of particles in the system. As we saw in the example of Eqs. \eqref{eq.EN.f} and \eqref{eq.EN.rho}, if a dependence of a functional on combinations of three or more particles can be neglected (as in case of energy), the average of the functional can be expressed in terms of only $\rho_1$ and $\rho_2$, which means that higher order distribution functions $\rho_3,\dots$ can be neglected. Transformation \eqref{eq.f.rho} can be seen as the transformation after which the total energy does not depend on $\rho_3$ and higher densities if there are only binary interactions present in the system. Transformation \eqref{eq.f.rho} thus makes it possible to express functionals in terms of only a few distribution functions.

However, neglecting higher order distribution functions has to be seen as an approximation which keeps precise values of energy (when considering only binary interactions) and number of particles. The situation is quite similar to that in BBGKY hierarchy, where it was shown in chapter 3.1.b of \cite{Hirschfelder} that macroscopic properties, namely the equation of state, are given by the first two distribution functions (when considering only binary interactions). That then leads to the assumption that non-equilibrium evolution is then also expressed just in terms of the first two distribution functions as in chapter 7.1.a of \cite{Hirschfelder}. In the grand-canonical ensemble energy is expressed only in terms of $\rho_1$ and $\rho_2$ and thus the equilibrium values of these two distribution functions are sufficient to reconstruct the fundamental thermodynamic relation $E=E(S,V,N)$. That means, as well as in \cite{Hirschfelder}, that it is reasonable to formulate also non-equilibrium evolution just in terms of the two first distribution functions. Moreover, the setting where only the first two distribution functions play the role of state variables is compatible with the pair-correlation formalism from \cite{Morito-Hiroike,deDominicus1962}.

On the other hand, neglecting higher-order distribution functions in the standard BBGKY hierarchy, or neglecting higher $e_n$ distribution functions, is rather problematic, since lower-order distribution functions $\ei$ are integrals of higher-order distribution functions $\ej$, $i < j$, and thus all the distribution functions would become zero. This is no longer true about the distribution functions $\rho_1, \rho_2, \dots$ in the GCE hierarchy, where higher-order distribution functions can be neglected without losing the lower-order distribution functions.

Formula \eqref{en.eq} together with the definition of $\rho_n$ give the equilibrium grand-canonical distribution functions,
\begin{equation}
 \rho_n|_{eq} = \frac{1}{\Xi}\frac{e^{-\beta\sum_{i=1}^n \frac{\pp_j^2}{2m}}\left(\frac{e^{\beta\mu}}{h^3}\right)^n}
 {V\left(\frac{2\pi m}{\beta}\right)^{3/2}}
 \exp\left(\frac{e^{\beta\mu}}{h^3} V \left(\frac{2\pi m}{\beta}\right)^{3/2}\right)
\end{equation}
where interactions $h_2$ are not taken into account for simplicity. Taking results from pages 83 and 84 of book \cite{Kubo}, it can be shown that
\begin{equation}
 \frac{\rho_n}{\intd (\mathbf{n}+\mathbf{1}) \rho_{n_1}} = \frac{1}{N},
\end{equation}
where $N$ is number of particles in the system. So in an ideal system (without interactions) distribution function $\rho_2$ can be neglected with respect to $\rho_1$. It can be expected that if binary but not ternary interactions are present, distribution function $\rho_3$ can be neglected while working only with $\rho_1$ and $\rho_2$, which is the case considered in this section.

Let us now turn to the evolution of the new distribution functions. Consider all functionals of the following form
\begin{equation}\label{eq.AAn}
 A = \sum_{n=1}^\infty A_n(\en)
\end{equation}
where each $A_n$ depends only on $\en$. Evolution of such distribution function $\en$ is given by Liouville equation \eqref{PB.eq.Liouville} because the prefactor in Eq. \eqref{eq.enfn} does not depend on the actual positions of the particles, i.e.
\begin{equation}
 \frac{\partial \en}{\partial t} = \en^0\left(-\frac{\partial H_n}{\partial p^i_\alpha} \frac{\partial \fn}{\partial r^\alpha_i} + \frac{\partial H_n}{\partial r^\alpha_i}\frac{\partial \fn}{\partial p^i_\alpha}\right),
\end{equation}
which implies that the functional $A_n$ evolves as
\begin{eqnarray}
 \frac{\partial A_n}{\partial t} &=& \intd\OO\dots\intd\nn \frac{\partial A_n}{\partial \en}\frac{\partial \en}{\partial t}=\nonumber\\
 &=&n! \underbrace{\intd\OO\dots\intd\nn \en \left(\frac{\partial}{\partial r^\alpha_i}\frac{\partial A_n}{\partial e_n}\frac{\partial}{\partial p^i_\alpha}\frac{\partial E_n}{\partial e_n}-\frac{\partial}{\partial r^\alpha_i}\frac{\partial E_n}{\partial e_n}\frac{\partial}{\partial p^i_\alpha}\frac{\partial A_n}{\partial e_n}\right)}_{\PB{A_n}{E_n}{GCEn}}.
\end{eqnarray}
Functional $A$ introduced in \eqref{eq.AAn} then evolves as
\begin{eqnarray}\label{eq.Aevo}
 \frac{\partial A}{\partial t} &=& \sum_{n=1}^\infty n! \PB{A_n}{E}{GCEn} = \sum_{n=1}^\infty n! \PB{A_n}{E_n}{GCEn}.
\end{eqnarray}
Note that the second equality follows from the fact that $\ei$ and $\ej$ are not coupled by the Poisson bracket when $i\neq j$.
Finally, we can introduce a grand-canonical Poisson bracket
\begin{equation}\label{eq.GCE.PB}
\PB{A}{B}{GCE} = \sum_{n=1}^\infty n! \PB{A}{B}{GCEn},
\end{equation}
which governs evolution of functionals of type \eqref{eq.AAn}, and Eq. \eqref{eq.Aevo} can be rewritten as
\begin{equation}
 \frac{\partial A}{\partial t} = \PB{A}{E}{GCE}.
\end{equation}
This bracket is indeed Poisson as r.h.s. of this last equation is again of form \eqref{eq.AAn} and as the bracket fulfills Jacobi identity (Liouville-Poisson brackets fulfill the identity for all $n$). Functionals of form \eqref{eq.AAn} thus form a Lie algebra with Poisson bracket \eqref{eq.GCE.PB}.

Moreover, restriction to functionals dependent only on $\en$, which is equivalent to projecting the set $(e_1,e_2,\dots)$ only to $\en$, yields the standard Liouville equation. Therefore, the algebra of functionals that depend  only on n-particle distribution functions with the Liouville-Poisson bracket is a projection from the algebra of grand-canonical ensemble functionals. The GCE hierarchy is thus a generalization of the hierarchy starting from the Liouville equation in that sense.

Poisson bracket \eqref{eq.GCE.PB} can be now expressed in terms of the new distribution functions $\rho_1$, $\rho_2$, $\dots$. Neglecting all distribution functions except for $\rho_1$ and $\rho_2$, the one-particle Liouville-Poisson bracket becomes
\begin{multline}
\PB{A(\rho_1,\rho_2,\dots)}{B(\rho_1,\rho_2,\dots)}{L1}=\\
=\intd\OO\left(\rho_1(\OO)-\intd\TT\rho_2(\OO,\TT)\right)\left(\frac{\partial}{\partial r^k_1}A_{\rho_1}(\OO)\frac{\partial}{\partial p^1_k} B_{\rho_1(\OO)}-\dots\right)
\end{multline}
where equations \eqref{eq.f.rho} and \eqref{eq.GCE.Afn} were used. Dots denote the same term as the preceding term inside the bracket but with $A$ and $B$ swapped. Similarly, the two-particle Liouville-Poisson bracket is
\begin{multline}
 \PB{A(\rho_1,\rho_2,\dots)}{B(\rho_1,\rho_2,\dots)}{L2}= \frac{1}{2}\intd\OO\intd\TT \rho_2(\OO,\TT)\left(\frac{\partial}{\partial r_1^k}A_{\rho_1(\OO)}\frac{\partial}{\partial p^1_k}B_{\rho_1(\OO)}-\dots\right)+\\
 + \frac{1}{2}\intd\OO\intd\TT \rho_2(\OO,\TT)\left(\frac{\partial}{\partial r_1^k}A_{\rho_1(\OO)}\frac{\partial}{\partial p^1_k}\left(B_{\rho_2(\OO,\TT)}+B_{\rho_2(\TT,\OO)}\right)-\dots\right)+\\
 + \frac{1}{2}\intd\OO\intd\TT \rho_2(\OO,\TT)\left(\frac{\partial}{\partial r_1^k}\left(A_{\rho_2(\OO,\TT)}+A_{\rho_2(\TT,\OO)}\right)\frac{\partial}{\partial p^1_k}\left(B_{\rho_1(\OO)}\right)-\dots\right)+\\
 + \frac{1}{2}\intd\OO\intd\TT \rho_2(\OO,\TT)\left(\frac{\partial}{\partial r_1^k}\left(A_{\rho_2(\OO,\TT)}+A_{\rho_2(\TT,\OO)}\right)\frac{\partial}{\partial p^1_k}\left(B_{\rho_2(\OO,\TT)}+B_{\rho_2(\TT,\OO)}\right)-\dots\right)
\end{multline}
Finally, bracket \eqref{eq.GCE.PB} becomes, when neglecting $\rho_3$ and higher distribution functions,
\begin{multline} \label{eq.GCE.PB.final}
  \PB{A(\rho_1,\rho_2,\dots)}{B(\rho_1,\rho_2,\dots)}{L<3}= \intd\OO\rho_1(\OO)\left(\frac{\partial}{\partial r^k_1}A_{\rho_1}(\OO)\frac{\partial}{\partial p^1_k} B_{\rho_1(\OO)}-\dots\right) +\\
 + \intd\OO\intd\TT \rho_2(\OO,\TT)\left(\frac{\partial}{\partial r_1^k}A_{\rho_1(\OO)}\frac{\partial}{\partial p^1_k}\left(B_{\rho_2(\OO,\TT)}+B_{\rho_2(\TT,\OO)}\right)-\dots\right)+\\
 + \intd\OO\intd\TT \rho_2(\OO,\TT)\left(\frac{\partial}{\partial r_1^k}\left(A_{\rho_2(\OO,\TT)}+A_{\rho_2(\TT,\OO)}\right)\frac{\partial}{\partial p^1_k}\left(B_{\rho_1(\OO)}\right)-\dots\right)+\\
 + \intd\OO\intd\TT \rho_2(\OO,\TT)\left(\frac{\partial}{\partial r_1^k}\left(A_{\rho_2(\OO,\TT)}+A_{\rho_2(\TT,\OO)}\right)\frac{\partial}{\partial p^1_k}\left(B_{\rho_2(\OO,\TT)}+B_{\rho_2(\TT,\OO)}\right)-\dots\right),
\end{multline}
which is surely a Poisson bracket as it can be seen as a transformation of Poisson bracket $\PB{A}{B}{L1}+\PB{A}{B}{L2}$ expressed in distribution functions $e_1$ and $e_2$ to distribution functions $\rho_1$ and $\rho_2$. Poisson bracket \eqref{eq.GCE.PB.final} expresses evolution of a two-point grand-canonical (GC) kinetic theory in a closed form, which enables us to describe the non-local character of fluid motion in the next section.

\subsection{From two-point grand-canonical kinetic theory to weakly non-local extended hydrodynamics}
Let us now consider a projection from the two-point kinetic theory to only a few moments of distributions $\rho_1$ and $\rho_2$.
\begin{subequations}\label{eq.tur.proj}
\begin{eqnarray}
 \rho(\rr) &=& \intd\pp m \rho_1(\rr,\pp)\\
 u_i(\rr) &=& \intd\pp p_i \rho_1(\rr,\pp)\\
 \label{eq.tur.s}s(\rr) &=& \intd\pp \sigma\left(\rho_1(\rr,\pp)-\intd\rr'\intd\pp'\rho_2(\rr,\pp,\rr',\pp')\right)\\
 b^{ij}(\rr)&=& \intd\RR\intd\pp\intd\PP R^iR^j \rhoh_2(\rr,\RR,\pp,\PP)\\
 c_{ij}(\rr)&=& \intd\RR\intd\pp\intd\PP P_iP_j \rhoh_2(\rr,\RR,\pp,\PP)\\
 w^i_j(\rr)&=& \intd\RR\intd\pp\intd\PP R^iP_j \rhoh_2(\rr,\RR,\pp,\PP)
\end{eqnarray}
\end{subequations}
where $\RR$, $\rr$, $\PP$ and $\pp$ are related to $\OO$ and $\TT$ by
\begin{subequations}
\begin{eqnarray}
 &&\rr = \frac{1}{2}(\rr_1+\rr_2),\qquad \pp = \pp^1+\pp^2,\\
 &&\RR = \rr_2-\rr_1,\qquad \PP = \frac{1}{2}(\pp^2-\pp^1),
\end{eqnarray}
and
\begin{equation}
 \rhoh_2(\rr,\RR,\pp,\PP) = \rho_2\left(\underbrace{\rr-\frac{\RR}{2}}_{\rr_1},\underbrace{\frac{\pp}{2}-\PP}_{\pp^1},\underbrace{\rr+\frac{\RR}{2}}_{\rr_2},\underbrace{\frac{\pp}{2}+\PP}_{\pp^2}\right).
\end{equation}
\end{subequations}
The symbol $\sigma$ stands for an arbitrary concave function from real numbers to real numbers. Projection \eqref{eq.tur.s} comes from expressing Eq. \eqref{eq.GCE.S} in terms of densities \eqref{eq.f.rho} when neglecting all densities except for $\rho_1$ and $\rho_2$. State variables \eqref{eq.tur.proj} consist of hydrodynamic state variables and three extra state variables expressing non-locality in position and velocity. We conjecture that these state variables could become useful when analyzing hydrodynamic turbulence because a very similar projection has been carried out in \cite{Miroslav-turbulence}, where the explicit relevance in turbulence modelling was demonstrated.

Derivatives of Eqs. \eqref{eq.tur.proj} with respect those densities are
\begin{subequations}
\begin{eqnarray}
 \frac{\partial \rho(\rr')}{\partial \rho_1(\OO)} &=& m\delta(\rr'-\rr_1)\\
 \frac{\partial u_i(\rr')}{\partial \rho_1(\OO)} &=& p^1_i \delta(\rr'-\rr_1)\\
 \frac{\partial s(\rr')}{\partial \rho_1(\OO)} &=& \delta(\rr'-\rr_1) \sigma'\left(\rho_1(\OO)-\intd\rr''\intd\pp''\rho_2(\rr_1,\pp^1,\rr'',\pp'')\right)\\
 \frac{\partial s(\rr')}{\partial \rho_2(\OO,\TT)} &=& -\frac{1}{2}\left(\sigma'\left(\rho_1(\OO)-\intd\rr''\intd\pp''\rho_2(\rr_1,\pp^1,\rr'',\pp'')\right)\delta(\rr_1-\rr')+\right.\nonumber\\
 &&+\left.\sigma'\left(\rho_1(\TT)-\intd\rr'''\intd\pp'''\rho_2(\rr_2,\pp^2,\rr''',\pp''')\right)\delta(\rr_2-\rr')\right)\nonumber\\
 &&\\
 \frac{\partial b^{ij}(\rr')}{\partial \rhoh_2(\rr,\RR,\pp,\PP)} &=& R^i R^j \delta(\rr'-\rr)\\
 \frac{\partial c_{ij}(\rr')}{\partial \rhoh_2(\rr,\RR,\pp,\PP)} &=& P_i P_j \delta(\rr'-\rr)\\
 \frac{\partial w^i_j(\rr')}{\partial \rhoh_2(\rr,\RR,\pp,\PP)} &=& R^i P_j \delta(\rr'-\rr).
\end{eqnarray}
\end{subequations}

Taking two arbitrary functionals of variables \eqref{eq.tur.proj}, expressing them in terms of $\rho_1$ and $\rho_2$ and plugging them into the Poisson bracket \eqref{eq.GCE.PB.final} leads (after relatively tedious calculations\footnote{available upon personal request to the corresponding author}) to an extended hydrodynamic Poisson bracket
\begin{multline}\label{PBturb}
 \PB{A(\rho,\uu,s,\bbb,\cc,\ww)}{B(\rho,\uu,s,\bbb,\cc,\ww)}{EH}=\PB{A}{B}{CH}+\PB{A}{B}{\mbox{conv}}+\\
 +\intd\rr b^{ij} ((A_{b^{kj}}+A_{b^{jk}})\partial_i B_{u_k}-(B_{b^{kj}}+B_{b^{jk}})\partial_i A_{u_k}) \\
 -\intd\rr c_{ij} ((A_{c_{kj}}+A_{c_{jk}})\partial_k B_{u_i}-(B_{c_{kj}}+B_{c_{jk}})\partial_k A_{u_i})\\
 +\intd\rr w^i_j (A_{w^k_j} \partial_i B_{u_k}-B_{w^k_j} \partial_i A_{u_k}) - w^j_i (A_{w^j_k}\partial_k B_{u_i}-B_{w^j_k}\partial_k A_{u_i})\\
 +2\intd\rr w^j_i\left(\left(A_{b^{kj}}+A_{b^{jk}}\right)\left(B_{c_{ki}}+B_{c_{ik}}\right)-\left(B_{b^{kj}}+B_{b^{jk}}\right)\left(A_{c_{ki}}+A_{c_{ik}}\right)\right)+\\
 +2\intd\rr b^{ji}\left(\left(A_{b^{kj}}+A_{b^{jk}}\right)B_{w^i_k}-\left(B_{b^{kj}}+B_{b^{jk}}\right)A_{w^i_k}\right)+\\
 +2\intd\rr c_{ji}\left(A_{w^k_j}\left(B_{c_{ki}}+B_{c_{ik}}\right)-B_{w^k_j}\left(A_{c_{ki}}+A_{c_{ik}}\right)\right)+\\
 +2\intd\rr w^i_j \left(A_{w^k_j}B_{w^i_k}-B_{w^k_j}A_{w^i_k}\right)
\end{multline}
where $\PB{A}{B}{CH}$ is the hydrodynamic Poisson bracket, Eq. \eqref{HB}, and $\PB{A}{B}{\mbox{conv}}$ is the convective part of the bracket,
\begin{eqnarray}
\PB{A}{B}{\mbox{conv}} &=&
 \intd\rr b^{ij}\left(\partial_k A_{b^{ij}} B_{u_k}-\partial_k B_{b^{ij}} A_{u_k}\right)\nonumber\\
 &&+\intd\rr c_{ij}\left(\partial_k A_{c_{ij}} B_{u_k}-\partial_k B_{c_{ij}} A_{u_k}\right)\nonumber\\
 &&+\intd\rr w^i_j\left(\partial_k A_{w^i_j} B_{u_k}-\partial_k B_{w^i_j} A_{u_k}\right).
\end{eqnarray}
The Poisson bracket (\ref{PBturb}) is closed within variables \eqref{eq.tur.proj} and thus generates the time evolution on the level of description given by the projection \eqref{eq.tur.proj}. Validity of the Jacobi identity has been checked using the automated tool developed in \cite{Kroeger2010}. The bracket can be regarded as a generalization\footnote{Poisson bracket \eqref{PBturb} contains also the entropy density field, and the factors $2$ were missing in \cite{Miroslav-turbulence}.} of the Poisson bracket from \cite{Miroslav-turbulence}.

When deriving this bracket we used a center-of-mass localization neglecting terms with second and higher spatial gradients of variational derivatives of the functionals with respect to the extended hydrodynamic fields \eqref{eq.tur.proj}. Such truncation of spatial Taylor expansions of the fields is used in terms multiplied by distribution function $\rho_2(\rr,\RR)$. If the distribution function decays rapidly with increasing particle distance $\RR$, it can be assumed that the extended hydrodynamic fields do not vary appreciably over such distance and that higher spatial derivatives of the fields can be neglected. Does the distribution function decay in $\RR$?

Our aim is to describe fluids with internal elastic degrees of freedom as polymeric fluids. Such elastic interaction energy can be written as
\begin{equation}
 \intd\pp\intd\PP\intd\rr\intd\RR \rho_2(\rr,\pp,\PP) h_2(\RR).
\end{equation}
To maintain finite value of energy, it can be thus expected that the two-particle distribution function $\rho_2$ decays with increasing $\RR$ rapidly. An estimate of the decay can be found in ES. Finally, assuming that the fields \eqref{eq.tur.proj} do not vary appreciably over the decay distance supports validity of the truncation of the spatial Taylor expansions.

In summary, two grand-canonical two-point kinetic theory developed in Sec. \ref{sec.GCE} can be projected to just a few moments of the distribution functions, namely the hydrodynamic moments and three weakly non-local tensors. Reversible evolution of these new state variables is then described by a Poisson bracket derived from the Poisson bracket of the two-point kinetic theory by projection.

\subsection{Spatial correlation function}\label{sec.g}
Another possibility is to project the first two distribution functions, $\rho_1$ and $\rho_2$, to the hydrodynamic fields, $\rho,\uu,s$, and a spatial correlation function
\begin{equation}
 g(\rr,\RR) = \intd\pp\intd\PP \hat{\rho}_2(\rr,\RR,\pp,\PP).
\end{equation}

Bracket \eqref{eq.GCE.PB.final} then reduces to
\begin{eqnarray}
 \PB{A}{B}{g} &=& \PB{A}{B}{CH} +\\
 &&+\intd\rr\intd\RR g(\rr,\RR)\frac{\partial}{\partial r^k}\left(\frac{A_g(\rr,\RR)+A_g(\rr,-\RR)}{2}\right)\cdot\nonumber\\
 &&\qquad\cdot \left(B_{u_k}(\rr)+\frac{1}{2!}\frac{R^l}{2}\frac{R^m}{2}\frac{\partial^2 B_{u_k}(\rr)}{\partial r^l\partial r^m}+\dots\right)-\dots \nonumber\\
 &&+\intd\rr\intd\RR g(\rr,\RR)R^l\frac{\partial}{\partial R^k}\left(\frac{A_g(\rr,\RR)+A_g(\rr,-\RR)}{2}\right)\cdot\nonumber\\
 &&\qquad\cdot \left(\frac{\partial B_{u_k}(\rr)}{\partial r^l}+\frac{1}{3!}\frac{R^l}{2}\frac{R^m}{2}\frac{R^n}{2}\frac{\partial^3 B_{u_k}(\rr)}{\partial r^l\partial r^m\partial r^n}+\dots\right)-\dots\nonumber
\end{eqnarray}
This bracket can be seen as a simplification, extension (by including also the entropy field) and correction (the factor $1/2$ and the higher spatial gradients) of the bracket introduced in \cite{MG-colloids2013}. The localization used when deriving Poisson bracket \eqref{PBturb} is not employed here as we do not assume that $\uu$ does not vary appreciably over the decay distance of $g$. When providing an appropriate energy functional $E(\rho,\uu,s,g)$, reversible part of evolution equations of the state variables can be obtained.

\section{Conclusion}
\begin{figure}[ht!]
\includegraphics{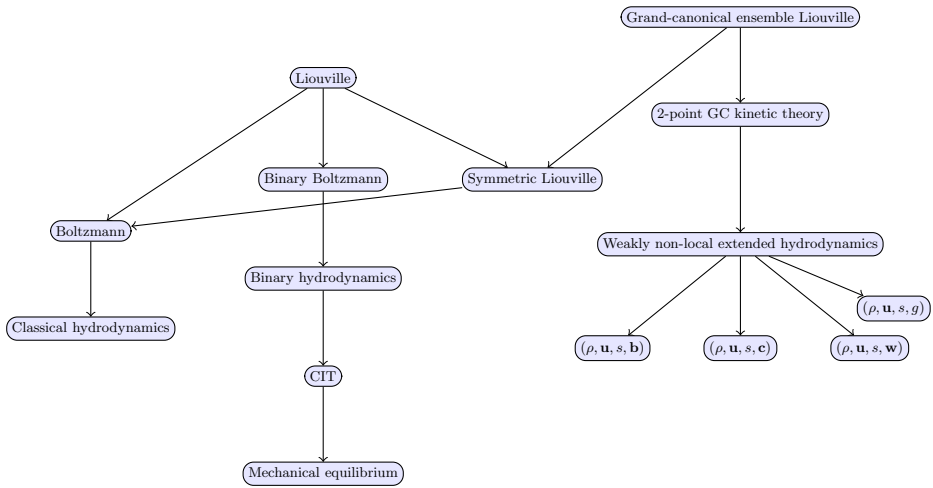}
\caption{Map of demonstrated projections. Note that if there is a projection from A to B and from B to C, then there is a projection from A to C. Starting from Liouville equation (or Liouville-Poisson bracket), Poisson brackets of Boltzmann equation and classical hydrodynamics are given by a straightforward projection. Similarly, Poisson bracket governing reversible evolution in kinetic theory of binary mixtures (binary Boltzmann equation) follows from the Liouville-Poisson bracket and it yields Poisson brackets for binary hydrodynamics (with two densities, momentum densities and entropy densities or temperatures) and for classical irreversible thermodynamics. Grand-canonical ensemble of Liouville equations, where number of particles can vary, can be simplified to the standard N-particle Liouville equation or it can be projected to a two-point kinetic theory, to weakly non-local extended hydrodynamics possibly useful in turbulence modelling or to the Poisson bracket coupling a spatial correlation function with hydrodynamic fields.}
\label{fig.map}
\end{figure}

The objective of this paper is to elucidate a method for projecting Poisson brackets expressing  kinematics on higher (more detailed) levels of description to  Poisson brackets expressing kinematics on lower (less detailed) levels in the context of non-equilibrium thermodynamics. The method is quite natural and widely used in differential geometry. Indeed, the new Poisson brackets  are obtained just by evaluating the original Poisson bracket on projections of the original state variables  onto  the lower level state variables. The pertinence and implications of the method (forming a hierarchy of Poisson brackets) in the physical context of non-equilibrium thermodynamics is discussed in Sec. \ref{sec.PB}.

This type of projections of Poisson bracket  is illustrated on two passages: (i) passage  from the Liouville-Poisson bracket  to the Boltzmann-Poisson bracket and subsequently to the hydrodynamic Poisson bracket, and (ii) passage  from the Liouville-Poisson bracket to the  Boltzmann-Poisson bracket for a binary mixture and subsequently  the Poisson bracket of the classical irreversible thermodynamics. The hierarchy of projections and Poisson brackets investigated in this paper is summarized in Fig. \ref{fig.map}.

Not all projections of state variables induce however projections of Poisson brackets.
For example,  the projection from the Liouville level to the levels that use  one-point and two-point distribution functions as state variables (the two-point part of the BBGKY hierarchy) leads to a Poisson bracket that  involves also    the three-point distribution function. In other words, the  bracket obtained after projection  is not   closed. In order to transform it into a bracket involving only the one-point and two-point distribution functions, we need to provide a closure, i.e. to express the three-point distribution function in terms of the one-and-two-point distribution functions.

Similarly,  the brackets arising in projections used to get the extended irreversible thermodynamics need closures. Is there a way how to ensure that the resulting bracket will depend only on state variables of the lower level (i.e. to ensure that we do not need to make any closures) before calculating the projection explicitly? The answer to this question is positive, but we have to know in addition the geometric structures of reversible dynamics on both more detailed and less detailed levels. For example, projection from the Liouville-Poisson bracket to the Boltzmann-Poisson bracket is ensured by the following geometrical argument. Firstly, reversible dynamics on both  levels can be regarded as generated by certain Lie groups. Secondly, Lie algebras of the Lie groups can be identified as N-particle and one-particle Hamiltonians which implies a natural homomorphism between those Lie algebras. Thirdly, the mapping dual to the homomorphism relates the N-particle distribution functions and the one-particle distribution functions. This mapping then preserves the Poisson bracket as it follows from differential geometry, e.g. \cite{Marsden-Ratiu}. Therefore, the resulting bracket is guaranteed to be a Poisson bracket. More details and examples are given in appendices of this paper.

The applicability of the projection method can be extended by considering
systems with variable contents, Sec. \ref{sec.GCE}. A grand-canonical ensemble of Liouville-Poisson brackets, allowing varying number of particles, is proposed. The grand-canonical ensemble is also viewed from the geometric perspective in \ref{sec.GCE.groups}. A two-point kinetic theory is  derived in a closed form by projection from the grand-canonical ensemble after physically motivated approximations. As an illustration, we investigate the  projection from the two-point kinetic theory to a weakly non-local extended hydrodynamics and identify the corresponding Poisson bracket. This particular application provides a setting that appears to be suitable for modeling of turbulent flows as the hydrodynamic fields are coupled with tensors expressing weakly non-local behavior, the conformation tensor, Reynolds stress tensor and non-local vorticity. A completely non-local Poisson bracket coupling a spatial correlation function with the hydrodynamic fields is derived in Sec. \ref{sec.g}.

In summary, we have shown in this paper that kinematics of reversible evolution (expressed in a Poisson bracket) can often be obtained  as a simple projection of the Poisson bracket expressing kinematics of the reversible time evolution on a higher level (more detailed) of description. One can for example start with Liouville equation and project the Liouville-Poisson bracket to many Poisson brackets describing evolution of less detailed descriptions, see Fig. \ref{fig.map}. When the calculation leads to a bracket in a closed form, the bracket is certainly Poisson and reversible in the sense of time-reversal \cite{PRE2014}. Moreover, even without carrying out the calculation, it is sometimes possible to ensure that the projection will result in a Poisson bracket by relating Lie algebras of the two levels of description. The hierarchy starting from the Liouville-Poisson bracket can be generalized to a grand-canonical hierarchy independent of number of particles, within which we obtain for example a two-point kinetic theory and a weakly non-local extended hydrodynamics, that appears to be suitable for modeling of turbulence, polymeric fluids or colloidal systems.

\section*{Acknowledgment}
We are grateful to professor František Maršík for his generous support and for revealing the world of thermodynamics to M. P. and V. K.

This project was supported by Natural Sciences and Engineering Research Council of Canada (NSERC).

The work was supported by Czech Science Foundation (project no. 14-18938S).

The work was partially developed within the POLYMEM project, reg. no CZ.1.07/2.3.00/20.0107, that is co-funded from the European Social Fund (ESF) in the Czech Republic: "Education for Competitiveness Operational Programme", from the CENTEM project, reg. no. CZ.1.05/2.1.00/03.0088, cofunded by the ERDF as part of the Ministry of Education, Youth and Sports OP RDI programme and, in the follow-up sustainability stage, supported through CENTEM PLUS (LO1402) by financial means from the Ministry of Education, Youth and Sports under the ”National Sustainability Programme I.``.

\appendix

\section{Geometry of the projection from Liouville to Boltzmann level}

\label{sec.LB.groups}

We have already demonstrated in Sec. \ref{PB.sec.B} that the Poisson bracket on the
Boltzmann level of description can be acquired by projection from the
Liouville level. But is it possible to see this possibility already from
symmetries of the two levels? Indeed, it is the goal of this section to
illustrate how symmetries imply that the projection defines a Poisson bracket on the
Boltzmann level of description.

We start with the inclusion
\begin{equation}
\psi :\mathfrak{g}\rightarrow \mathfrak{g}_{N}^{s}:\mathcal{F}\left( T^{\ast
}Q\right) \rightarrow \mathcal{F}\left( T^{\ast }M_{N}\right) :h\rightarrow
\psi h=\sum_{i=1}^{N}h\left( \mathbf{i}\right) .  \label{iso}
\end{equation}%
from the algebra $\mathfrak{g}$ of functions on $T^{\ast }Q$ to the algebra $%
\mathfrak{g}_{N}^{s}$ of symmetric functions on $T^{\ast }M_{N}=T^{\ast
}\left( Q_{1}\times ...\times Q_{N}\right)$. The mapping $\psi$ in \eqref{iso} is a Lie algebra homomorphism, which means that for two arbitrary functions $h$ and $k$ on $T^{\ast }Q$, the
identity
\begin{equation*}
\left\{ \psi h,\psi k\right\} ^{\left( CM\right) }=\psi \left\{ h,k\right\}
^{\left( CM\right) }
\end{equation*}%
holds, see e.g. \cite{Marsden-BBGKY}. Linear algebraic dual
\begin{equation}
\psi ^{\ast }:\mathfrak{g}_{N}^{\ast }\rightarrow \mathfrak{g}^{\ast }:f_{N}\left(
\mathbf{1,...,N}\right) d\mathbf{1}.\mathbf{..}d\mathbf{N}\rightarrow
f\left( \mathbf{a}\right) d\mathbf{a}  \label{Pois}
\end{equation}%
of $\psi $ is from the space of densities $\mathfrak{g}_{N}^{\ast }$ on $%
T^{\ast }M_{N}$ to the space of densities $\mathfrak{g}^{\ast }$ on $T^{\ast
}Q$ and according to the calculation
\begin{eqnarray*}
&&\left\langle \psi h\left( \mathbf{1,...,N}\right),f_{N}\left( \mathbf{1}%
,...,\mathbf{N}\right) d\mathbf{1}...d\mathbf{N}\right\rangle _{T^{\ast
}M_{N}}\\&&=
\int d\mathbf{1...}\int d\mathbf{N}\psi h\left( \mathbf{1,...,N}%
\right) f_{N}\left( \mathbf{1,...,N}\right)  \\&&
=\int d\mathbf{1...}\int d\mathbf{N}\left( h\left( \mathbf{1}\right)
+...+h\left( \mathbf{N}\right) \right) f_{N}\left( \mathbf{1,...,N}\right)
\\&&
=\int d\mathbf{a\int }d\mathbf{\mathbf{2}...}\int d\mathbf{N}h\left(
\mathbf{a}\right) f_{N}\left( \mathbf{a,2,...,N}\right)  \\&&
+...+\mathbf{\int }d\mathbf{1...}\int d\left( \mathbf{N-1}\right) \int d%
\mathbf{a}h\left( \mathbf{a}\right) f_{N}\left( \mathbf{1,...,N-1,a}\right)\\&&
=\int d\mathbf{a}h\left( \mathbf{a}\right) \int d\mathbf{1...}\int d%
\mathbf{N}f_{N}\left( \mathbf{1,...,N}\right) \sum_{i=1}^{N}\delta (\mathbf{a%
}-\mathbf{i}),  \label{calc}
\end{eqnarray*}%
explicitly given by%
\begin{equation}
f\left( \mathbf{a}\right) =\int d\mathbf{1...}\int d\mathbf{N}%
f_{N}\sum_{i=1}^{N}\delta (\mathbf{a}-\mathbf{i}).  \label{f}
\end{equation}%
Note that, what we derived in \eqref{f} is exactly the definition in \eqref{PB.eq.ffN}. Dual of Lie algebra
homomorphism is a momentum and Poisson map, see Proposition 10.7.2 in \cite%
{Marsden-Ratiu}. For our case, this means that the functionals on $\mathfrak{%
g}_{N}^{\ast }$ have the same Poisson bracket as functionals on $\mathfrak{g}%
^{\ast }$. In other words, the identity \eqref{PB.eq.LB} holds since the projection $\psi ^{\ast }$ is dual of a Lie algebra homomorphism.

In order to see how the Poisson mapping $\psi ^{\ast }$ in \eqref{Pois}
explicitly works, we consider two linear functionals
\begin{equation*}
A\left( fd\mathbf{a}\right) =\int_{T^{\ast }Q}d\mathbf{a}\, h\left( \mathbf{a}%
\right) f\left( \mathbf{a}\right) \text{ \ \ and \ \ }B\left( fd\mathbf{a}%
\right) =\int_{T^{\ast }Q}d\mathbf{a}\, k\left( \mathbf{a}\right) f\left(
\mathbf{a}\right)
\end{equation*}%
on the space $\mathfrak{g}^{\ast }$ of densities on $T^{\ast }Q$. Here, $h$
and $k$ are a real-valued functions on $T^{\ast }Q$. Under the reflexivity
condition $\mathfrak{g}^{\ast \ast }\simeq \mathfrak{g}$, the
functional derivatives of $A$ and $B$ with respect to their arguments result
with
\begin{equation*}
\frac{\partial A}{\partial f}=h\left( \mathbf{a}\right) \text{ \ \ and \ \ }%
\frac{\partial B}{\partial f}=k\left( \mathbf{a}\right) ,
\end{equation*}%
respectively. Pullbacks of the functionals $A$ and $B$ by the Poisson
mapping $\psi ^{\ast }$ in \eqref{Pois}\ are linear functionals on the space $%
\mathfrak{g}_{N}^{\ast }$ of densities on $T^{\ast }\left( M_{N}\right) $
given by for example
\begin{equation}
A\left( f_{N}\right) =\int d\mathbf{1}\int d\mathbf{2...}\int d\mathbf{N}%
h_{N}f_{N}  \label{B}
\end{equation}%
whereas the functional derivatives of $A$ is
\begin{equation*}
\frac{\partial A}{\partial f_{N}}=h_{N}\left( \mathbf{1,2,...,N}\right)
=h\left( \mathbf{1}\right) +h\left( \mathbf{2}\right) +...+h\left( \mathbf{N}%
\right) .
\end{equation*}%
By the abuse of notation, we are using the same notation $A$ both for the
functionals $A$ on $\mathfrak{g}^{\ast }$ and $\mathfrak{g}_{N}^{\ast }$. We
write Liouville Poisson bracket as follows
\begin{eqnarray*}
&&\left\{ A,B\right\} ^{\left( L\right) }\left( f_{N}\right)=\int d\mathbf{1...}\int d\mathbf{N}f_{N}\left( \mathbf{1,...,N}\right)
\left\{ \frac{\partial A}{\partial f_{N}},\frac{\partial B}{\partial f_{N}}%
\right\} ^{\left( CM\right) } \\
&=&\int d\mathbf{1...}\int d\mathbf{N}f_{N}\left( \mathbf{%
1,...,N}\right) \left\{ h_{N}\left( \mathbf{1,...,N}\right) ,k_{N}\left(
\mathbf{1,...,N}\right) \right\} ^{\left( CM\right) } \\
&=&\int d\mathbf{1...}\int d\mathbf{N}f_{N}\left( \mathbf{%
1,...,N}\right) \left\{ h\left( \mathbf{1}\right) +...+h\left( \mathbf{N}%
\right) ,k\left( \mathbf{1}\right) +...+k\left( \mathbf{N}\right) \right\}
^{\left( CM\right) } \\
&=&\int d\mathbf{1...}\int d\mathbf{N}f_{N}\left( \mathbf{%
1,...,N}\right) \left\{ h\left( \mathbf{1}\right) ,k\left( \mathbf{1}%
\right) \right\} ^{\left( CM\right) } \\
&&+...+\int d\mathbf{1...}\int d\mathbf{N}f_{N}\left(
\mathbf{1,...,N}\right) \left\{ h\left( \mathbf{N}\right) ,k\left( \mathbf{%
N}\right) \right\} ^{\left( CM\right) } \\
&=&\int d\mathbf{1...}\int d\mathbf{N}f_{N}\left( \mathbf{%
1,...,N}\right) \left\{ h\left( \mathbf{a}\right) ,k\left( \mathbf{a}%
\right) \right\} ^{\left( CM\right) }\sum_{i=1}^{N}\delta (\mathbf{a}-%
\mathbf{i}) \\
&=&\int d\mathbf{a}f\left( \mathbf{a}\right) \left\{ h\left( \mathbf{a}%
\right) ,k\left( \mathbf{a}\right) \right\} ^{\left( CM\right) }
=\left\{ A,B\right\} ^{\left( B\right) }\left( f\right),
\end{eqnarray*}%
where we used the Poisson mapping \eqref{f} and the definition
of the Boltzmann Poisson bracket on $\mathfrak{g}^{\ast }$ in the last line.
Therefore, in a theoretical level \eqref{PB.eq.LB} is established.

Let us end this section by writing some comments about equation \eqref%
{PB.eq.LBLL} relating Poisson bivectors. In the setting described in this
section, the Poisson bivector $L_{B}$ defined in \eqref{PB-B}\ for the
Boltzmann Poisson bracket is the Hamiltonian operator
\begin{equation*}
L_{B}:\mathfrak{g}^{\ast }\rightarrow \left( \mathfrak{g}\rightarrow
\mathfrak{g}\right) :f\rightarrow L_{B}\left( f\right) =X_{f}
\end{equation*}%
mapping a density function $f\in \mathfrak{g}^{\ast }$ to a linear
differential operator $L_{B}\left( f\right) $ on $\mathfrak{g}$, see \cite{Ol93}%
. A direct calculation shows that value of the operator $L_{B}\left(
f\right) $ on an element $h\in \mathfrak{g}$ is simply the directional
derivative of $h$ in the direction of Hamiltonian vector field $X_{f}$.
Similarly, Poisson bivector $L_{L}$ corresponding to the Liouville
Poisson bracket can be considered as the Hamiltonian operator
\begin{equation*}
L_{L}:\mathfrak{g}_{N}^{\ast }\rightarrow \left( \mathfrak{g}_{N}\rightarrow
\mathfrak{g}_{N}\right) :f_{N}\rightarrow L_{B}\left( f_{N}\right) =X_{f_{N}}.
\end{equation*}
The passage from the Liouville operator $L_{L}$ to Boltzmann operator $L_{B}$
is%
\begin{equation}
L_{L}\left( f_{N}\right) \circ \psi =L_{B}\left( \psi ^{\ast }\circ
f_{N}\right) ,  \label{HOI}
\end{equation}%
where $\psi $ is the Lie algebra homomorphism in \eqref{iso} and $\psi ^{\ast }$
is the dual of $\psi $ given in \eqref{Pois}. To prove \eqref{HOI}, we compute
the following:
\begin{eqnarray*}
&&\int d\mathbf{1...}\int d\mathbf{N} L_{L} \left( f_{N}\right) \circ \psi
\left( k\right) =\int d\mathbf{1...}\int d\mathbf{N}X_{f_{N}}\left( \psi
k\right) \\ &=&\int d\mathbf{1...}\int d\mathbf{N}X_{f_{N}}\left( k\left( \mathbf{%
1}\right) +...+k\left( \mathbf{N}\right) \right) \\
&=&\int d\mathbf{a...}\int d\mathbf{N}X_{f_{N}}k\left( \mathbf{a}\right)
+...+\int d\mathbf{1...}\int d\left( \mathbf{N-1}\right) \int d\mathbf{a}%
X_{f_{N}}k\left( \mathbf{a}\right) \\
&=&( \int d\mathbf{a...}\int d\mathbf{N}+...+\int d\mathbf{1...}\int
d\left( \mathbf{N-1}\right) \int d\mathbf{a}) \frac{\partial f_{N}}{%
\partial p_{a}}\frac{\partial k\left( \mathbf{a}\right) }{\partial r^{a}}-%
\frac{\partial f_{N}}{\partial r^{a}}\frac{\partial k\left( \mathbf{a}%
\right) }{\partial p_{a}} \\
&=&\int d\mathbf{a}\frac{\partial k\left( \mathbf{a}\right) }{\partial r^{a}}%
\frac{\partial }{\partial p_{a}}\left( \int d\mathbf{1...}\int d\mathbf{N}%
f_{N}\sum_{i=1}^{N}\delta (\mathbf{a}-\mathbf{i})\right) \\&&-\int d\mathbf{a}%
\frac{\partial k\left( \mathbf{a}\right) }{\partial p_{a}}\frac{\partial }{%
\partial r^{a}}\left( \int d\mathbf{1...}\int d\mathbf{N}f_{N}\sum_{i=1}^{N}%
\delta (\mathbf{a}-\mathbf{i})\right) \\
&=&\int d\mathbf{a}\frac{\partial k\left( \mathbf{a}\right) }{\partial r^{a}}%
\frac{\partial f\left( \mathbf{a}\right) }{\partial p_{a}}-\int d\mathbf{a}%
\frac{\partial k\left( \mathbf{a}\right) }{\partial p_{a}}\frac{\partial
f\left( \mathbf{a}\right) }{\partial r^{a}}=\int d\mathbf{a}L_{B}\left(
f\right) k.
\end{eqnarray*}

In summary, the dual $\mathfrak{g}_{N}^{\ast }\rightarrow \mathfrak{g}^{\ast }$ of the Lie algebra
embedding $\mathfrak{g}\rightarrow \mathfrak{g}_{N}^{s}$ in \eqref{iso}
is the Poisson mapping in \eqref{PB.eq.ffN} which maps the functionals
on $\mathfrak{g}_{N}^{\ast }$ to the functionals on $\mathfrak{g}^{\ast }$
while preserving the Poisson structure as described in \eqref{PB.eq.LB}. This equivalence is established not only in the level of Poisson brackets but also in the level of Poisson bivectors which is summarized in \eqref{HOI}.
\begin{remark}
The following question may arise at this point. Is it possible to write the set of projections from the Boltzmann Poisson bivector $L_{B}$ to the classical hydrodynamics Poisson bracket $L_{CH}$ as dual of a Lie algebra homomorphism? The projections to $\rho$ and $\uu$, the first two moments of the distribution function $f$ on $T^{\ast}Q$, are Poisson mappings \cite{Marsden-Ratiu}. To add the entropy into this picture is not possible in a straightforward way, since the dependence of entropy on the distribution function is nonlinear.
\end{remark}

\section{Geometry of the Grand-canonical Ensemble}

\label{sec.GCE.groups}

In this section, we construct a geometric framework for the grand-canonical
Poisson bracket \eqref{eq.GCE.PB} and derive both the families of
distribution functions \eqref{eq.f.rho} and \eqref{eq.rho.f} as duals of some injective
Lie algebra homomorphisms. As summarized in the previous sections, duals of
Lie algebra homomorphisms are necessarily Poisson \cite{Marsden-Ratiu}. For GCE
case, deriving \eqref{eq.f.rho} as a Poisson map will verify the calculations
done in \eqref{eq.GCE.PB.final} in a geometrical level. The motivation of this
section and some basic definitions comes from \cite{Marsden-BBGKY} where the
Lie-Poisson picture of BBGKY hierarchy was established. We assume that all
functional analytic conditions are satisfied.

For $N>M$, consider the following mapping
\begin{equation}
\mathfrak{g}_{M}^{s}\rightarrow \mathfrak{g}_{N}^{s}:h_{M}\rightarrow \sum_{
_{\substack{ i_{1},...,i_{M}=1  \\ i_{1}\neq ...\neq i_{M}}}}^{N}\frac{1}{M!}%
h_{M}\left( \mathbf{i}_{1}\mathbf{,...},\mathbf{i}_{M}\right)  \label{nN}
\end{equation}%
from the Lie algebra of the symmetric functions $\mathfrak{g}_{M}^{s}=%
\mathcal{F}\left( T^{\ast }M_{M}\right) $ into the Lie algebra $\mathfrak{g}%
_{N}^{s}=\mathcal{F}\left( T^{\ast }M_{N}\right) $ as in \cite{Marsden-BBGKY}. Note
that, when $M=1$, this definition reduces to the one in \eqref{iso}.

We denote the
infinite direct sum of algebras of the symmetric functions $\mathfrak{g}%
_{N}^{s}=\mathcal{F}\left( T^{\ast }M_{N}\right) $ by $\bigoplus_{N=1}^{%
\infty }\mathfrak{g}_{N}^{s}$. For all $N$ we embed the spaces $\mathfrak{g}%
_{M}^{s}$ into the infinite product as follows:
\begin{eqnarray}\label{Lal}
\psi _{1} &:&\mathfrak{g}\rightarrow \bigoplus_{N=1}^{\infty }\mathfrak{g}%
_{N}^{s}:h_{1}\rightarrow \bigoplus_{N=1}^{\infty }\sum_{i=1}^{N}h_{1}\left(
\mathbf{i}\right) =( h_{1}\left( \mathbf{1}\right) ,h_{1}\left( \mathbf{%
1}\right) +h_{1}\left( \mathbf{2}\right) ,...,\sum_{i=1}^{N}h_{1}\left(
\mathbf{i}\right) ,...)  \notag \\
\psi _{2} &:&\mathfrak{g}_{2}^{s}\rightarrow \bigoplus_{N=1}^{\infty }%
\mathfrak{g}_{N}^{s}:h_{2}\rightarrow \bigoplus_{N=1}^{\infty }\sum
_{\substack{ i_{1},i_{2}=1  \\ i_{1}\neq i_{2}}}^{N}\frac{1}{2!}h_{2}\left(
\mathbf{i}_{1}\mathbf{,i}_{2}\right) =( 0,h_{2}\left( \mathbf{1,2}%
\right) ,...,\sum_{_{\substack{ i_{1},i_{2}=1  \\ i_{1}\neq i_{2}}}}^{N}%
\frac{1}{2!}h_{2}\left( \mathbf{i}_{1}\mathbf{,i}_{2}\right) ,...)
\notag \\
&&...  \notag \\
\psi _{M} &:&\mathfrak{g}_{M}^{s}\rightarrow \bigoplus_{N=1}^{\infty }%
\mathfrak{g}_{N}^{s}:h_{M}\rightarrow \bigoplus_{N=1}^{\infty }\sum_{
_{\substack{ i_{1},...,i_{M}=1  \\ i_{1}\neq ...\neq i_{M}}}}^{N}\nonumber\\
&&\frac{1}{M!}%
h_{M}\left( \mathbf{i}_{1}\mathbf{,...},\mathbf{i}_{M}\right) =(
0,...,h_{M}\left( \mathbf{1,...},\mathbf{M}\right) ,...,\sum_{_{\substack{ %
i_{1},...,i_{M}=\nonumber\\
=1  \\ i_{1}\neq ...\neq i_{M}}}}^{N}\frac{1}{M!}h_{M}\left(
\mathbf{i}_{1}\mathbf{,...},\mathbf{i}_{M}\right) ,...)  \notag \\
&&....
\end{eqnarray}

An element $\mathfrak{H\in }\bigoplus_{N=1}^{\infty }\mathfrak{g}_{N}^{s}$
can be represented by an infinite sequence $\left(
h_{1},h_{2},...,h_{N},...\right) $. Taking the direct product of the domains
of the family of the mappings $\psi _{M}$ in (\ref{Lal}) leads to the mapping%
\begin{equation*}
\hat{\psi}:\bigoplus_{n=1}^{\infty }\mathfrak{g}_{n}^{s}\rightarrow
\bigoplus_{n=1}^{\infty }\mathfrak{g}_{n}^{s}:\bigoplus_{N=1}^{\infty
}h_{n}\rightarrow \hat{\psi}\left( \mathfrak{H}\right) =\sum_{n=1}^{\infty
}\psi _{n}\left( h_{n}\right),
\end{equation*}%
where on the right hand side we perform the addition term by term
\begin{eqnarray}
\hat{\psi}\left( \mathfrak{H}\right) =\sum_{i=1}^{\infty }\psi _{i}\left(
h_{i}\right) =( h_{1}\left( \mathbf{1}\right)
,...,\sum_{i=1}^{N}h_{1}\left( \mathbf{i}\right) +\frac{1}{2!}\sum
_{\substack{ i_{1},i_{2}=1 \\ i_{1}\neq i_{2}}}^{N}h_{2}\left( \mathbf{i}_{1}%
\mathbf{,i}_{2}\right) +...\notag\\+\sum_{_{\substack{ i_{1},...,i_{M}=1 \\ %
i_{1}\neq ...\neq i_{M}}}}^{N}\frac{1}{M!}h_{M}\left( \mathbf{i}_{1}\mathbf{%
,...},\mathbf{i}_{M}\right) ,...) .  \label{psi-hat}
\end{eqnarray}%
Consider two elements $\mathfrak{H}$ and $\mathfrak{K}$ of the product space
given by infinite sequences $\mathfrak{H}=\left( h_{1},...,h_{N},...\right) $
and $\mathfrak{K}=\left( k_{1},...,k_{N},...\right) $. In the domain of $%
\hat{\psi}$, define the following Lie algebra structure
\begin{equation} \label{LApro}
\left\{ \mathfrak{H},\mathfrak{K}\right\} =\bigoplus_{n+m=2}^{\infty
}\left\{ h_{n}\left( \mathbf{i}_{1}\mathbf{,...},\mathbf{i}_{n}\right)
,k_{m}\left( \mathbf{j}_{1}\mathbf{,...},\mathbf{j}_{m}\right) \right\},
\end{equation}%
which is the infinite version of the one introduced in \cite{Marsden-BBGKY}.
In the image space, consider the Lie algebra structure
\begin{equation*} \label{LApro2}
\left\{ \mathfrak{H},\mathfrak{K}\right\} =\bigoplus_{\substack{ n+m=2 \\ r+s=n+m-1}}^{\infty
}\left\{ h_{n}\left( \mathbf{i}_{1}\mathbf{,...},\mathbf{i}_{n}\right)
,k_{m}\left( \mathbf{j}_{1}\mathbf{,...},\mathbf{j}_{m}\right) \right\}-
\left\{ h_{r}\left( \mathbf{i}_{1}\mathbf{,...},\mathbf{i}_{r}\right)
,k_{s}\left( \mathbf{j}_{1}\mathbf{,...},\mathbf{j}_{s}\right) \right\}.
\end{equation*}
which is explicitly in form
\begin{eqnarray*}
\left\{ \mathfrak{H},\mathfrak{K}\right\} &=&(\left\{h_{1},k_{1}\right\},\left\{h_{1},k_{2}\right\}+\left\{h_{2},k_{1}\right\}-\left\{h_{1},k_{1}\right\},
\\&&\left\{h_{1},k_{3}\right\}+\left\{h_{2},k_{2}\right\}+\left\{h_{3},
k_{1}\right\}-\left\{h_{1},k_{2}\right\}-\left\{h_{2},k_{1}\right\},...).
\end{eqnarray*}
Note that the domain and image spaces are the same as
vector spaces but they have different Lie algebra structures. In this
construction, $\hat{\psi}$ becomes a Lie algebra homomorphism. In addition, $%
\hat{\psi}$ is injective hence has an inverse. The inverse mapping is defined as follows:
\begin{eqnarray}
h_{1} &=&k_{1},\text{ }h_{2}\left( 1,2\right) =k_{2}\left( 1,2\right)
-k_{1}\left( 1\right) -k_1\left( 2\right) ,...,
\end{eqnarray}
or, generally by the following formula
\begin{eqnarray}
h_{N}\left( \mathbf{1,...,N}\right) =\sum_{_{M=0}}^{N}\left( -1\right) ^{M}\sum_{_{\substack{ i_{1},...,i_{N-M}=1
\\ i_{1}\neq ...\neq i_{N-M}}}}^{N}\frac{1}{\left( N-M\right) !}%
k_{N-M}\left( \mathbf{i}_{1}\mathbf{,...},\mathbf{i}_{N-M}\right).  \label{psi-hat-inv}
\end{eqnarray}%
By definition, this inverse mapping is also a Lie algebra
homomorphism. Before proceeding the Lie-Poisson picture of the grand-canonical Poisson bracket, this may be a good point to make some comments about the projection from Liouville Poisson bracket to two-point BBGKY presented in Sec. \ref{Lto2BBGKY}.
\begin{remark} \label{Remark.L.2BBGKY}
In \cite{Marsden-BBGKY}, Poisson picture of the BBGKY hierarchy is obtained as the dualization of the Lie algebra homomorphism
\begin{equation*}
\bigoplus_{n=1}^{N}\mathfrak{g}_{n}^{s}\rightarrow\mathfrak{g}_{N}^{s}:\bigoplus_{n=1}^{N}h_{n}\rightarrow \sum_{n=1}^{N
}h_{n}.
\end{equation*}%
Here, in the domain, we define a finite version of the Lie algebra bracket \eqref{LApro}. For $M<N$, to guarantee that the projection from the Liouville Poisson bracket to $M$-point BBGKY be a Poisson map, one needs to embed the space of
one-two...-$M$-point functions $\bigoplus_{n=1}^{M}\mathfrak{g}_{n}^{s}$ into the total space $\bigoplus_{n=1}^{N}\mathfrak{g}_{n}^{s}$ as a Lie subalgebra. A direct calculation shows that, the only proper subalgebra of $\bigoplus_{n=1}^{N}\mathfrak{g}_{n}^{s}$ is $\mathfrak{g}_{1}$, which is the case $M=1$ consists of one-point functions. Embedding of the one-point functions into the total space is presented in \eqref{iso}, which is the main tool used in the \ref{sec.LB.groups}. The space of one-and-two point functions $\mathfrak{g}_{1}\oplus\mathfrak{g}_{2}^{s}$ can not be embedded into to total space $\bigoplus_{n=1}^{N}\mathfrak{g}_{n}^{s}$ as a subalgebra, so that the projection from the Liouville level to the two-point BBGKY can not be Poisson. This leads to the existence of the 3-particle distribution function $f_{3}$ in the calculation the Poisson bracket, which does prevent the bracket being closed. See ES for more details.
\end{remark}

Let us now proceed to the grand-canonical ensemble. We take the dual of the product Lie algebra as direct product of the dual
spaces $\mathfrak{g}_{n}^{s\ast }$ of $\mathfrak{g}_{n}^{s}$, that is
\begin{equation*}
\left( \bigoplus_{N=1}^{\infty }\mathfrak{g}_{N}^{s}\right) ^{\ast
}=\bigoplus_{N=1}^{\infty }\mathfrak{g}_{N}^{s\ast }.
\end{equation*}%
An element of the total space $\bigoplus_{N=1}^{\infty }\mathfrak{g}%
_{N}^{s\ast }$ is an infinite sequence
\begin{equation}
\mathfrak{e}=\left( e_{1}d\mathbf{1},e_{2}d\mathbf{1}d\mathbf{2},...,e_{N}d%
\mathbf{1}d\mathbf{2}...d\mathbf{N},... .\right) \label{en}
\end{equation}%
where the density functions $e_{N}$ are defined by multiplying each density
function $f_{N}\in \mathfrak{g}_{N}^{s\ast }$ with a scalar $e_{N}^{0}$ in
order to satisfy the normalization condition \eqref{norm-e}. The volume forms can be chosen particularly as
symplectic volumes $%
d\mathbf{1}d\mathbf{2}...d\mathbf{N}$. We define the duality between $\mathfrak{e}\in
\bigoplus_{N=1}^{\infty }\mathfrak{g}_{N}^{s\ast }$ and $\mathfrak{H}\in
\bigoplus_{N=1}^{\infty }\mathfrak{g}_{N}$ as the sum of dualizations of the
corresponding components multiplied by $1/N!$,
that is
\begin{equation*}
\left\langle \mathfrak{e},\mathfrak{H}\right\rangle =\sum_{N=1}^{\infty }%
\frac{1}{N!}\int ...\int d\mathbf{1}d\mathbf{2}...d\mathbf{N}e_{N}h_{N}.
\end{equation*}%
The Lie-Poisson bracket on $\bigoplus_{N=1}^{\infty }\mathfrak{g}_{N}^{s\ast
}$ is given in \eqref{eq.GCE.PB}. The dual of the Lie algebra isomorphism \eqref%
{psi-hat} is
\begin{equation*}
\hat{\psi}^{\ast }:\bigoplus_{N=1}^{\infty }\mathfrak{g}_{N}^{s\ast
}\rightarrow \bigoplus_{N=1}^{\infty }\mathfrak{g}_{N}^{s\ast }:\left( e_{1}d%
\mathbf{1},...,e_{N}d\mathbf{1}...d\mathbf{N},...\right)
\rightarrow \left( \rho _{1}d\mathbf{1},...,\rho _{N}d\mathbf{1}%
...d\mathbf{N},...\right)
\end{equation*}%
whose image space consists of the densities in form
\begin{equation}
\rho _{M}=\sum_{N=M}^{\infty }\frac{1}{\left( N-M\right) !}\int ...\int
d\left( \mathbf{M+1}\right) ...d\mathbf{N}e_{N}.  \label{rho-gen}
\end{equation}%
Here the first two elements $\rho _{1}$ and $\rho _{2}$ are as in \eqref{eq.f.rho}.

In summary, the new set of distribution functions $\rho_{M}$ in \eqref{rho-gen} is derived from the canonical ones $e_{N}$ in \eqref{en} as dual of the injective Lie algebra homomorphism $\hat{\psi}$ in \eqref{psi-hat}. This means that distribution functions \eqref{rho-gen} are components of a Poisson mapping relating the Poisson bracket in \eqref{eq.GCE.PB} to the Poisson bracket in \eqref{eq.GCE.PB.final}.

In other words, the way the total energy depends on the n-particle Hamiltonians implies the construction of mapping $\hat{\psi}$, dual of which maps the original grand-canonical distribution functions to the new distribution functions $\rho_1, \rho_2, \dots$. The construction of the grand-canonical ensemble itself thus already leads to the construction of the new distribution functions in a geometric way.

\end{document}